\documentclass[aps,twocolumn,showpacs,amsmath,amssymb,prc]{revtex4-2}
\usepackage{graphicx,here}
\usepackage{multirow}
\usepackage{tikz}
\usepackage{epstopdf}
\usepackage[percent]{overpic}
\usepackage{scrextend}
\usepackage{xcolor,xspace}
\usepackage[ulem=normalem]{changes}
\usepackage{hyperref}
\hypersetup{colorlinks=True,urlcolor=blue,linkcolor=blue,citecolor=blue,filecolor=black}

\colorlet{Changes@Color}{red}
\usepackage{dcolumn,color,footnote,bm,braket}
\usepackage{url,longtable,tabularx}
\usepackage{threeparttable}

\usepackage{fancybox}
\usetikzlibrary{matrix}

\newcommand\+{\dagger}
\newcommand\jr{j_{\rho}}
\newcommand\mr{m_{\rho}}

\newcommand\mgt{M^{\mathrm{GT}}}

\newcommand\bgtbeta{B({\mathrm{GT}})}

\newcommand\hb{\hat{H}_{\mathrm{B}}}
\newcommand\hf{\hat{H}_{\mathrm{F}}}
\newcommand\hbf{\hat{V}_{\mathrm{BF}}}
\newcommand\hff{\hat{V}_{\nu\pi}}

\newcommand\tnbb{2\nu\beta\beta}
\newcommand\znbb{0\nu\beta\beta}

\newcommand\ga{g_{\mathrm{A}}}
\newcommand\gv{g_{\mathrm{V}}}

\newcommand\ft{\log{}_{10}ft}
\newcommand\btm{\beta^{-}}

\newcommand\vd{v_{\mathrm{d}}}
\newcommand\vssd{v_{\mathrm{ssd}}}
\newcommand\vsss{v_{\mathrm{ss}}}
\newcommand\vt{v_{\mathrm{t}}}

\begin{document}

\title{Low-energy structure and $\beta$ decay properties of neutron-rich nuclei in the region of a shape phase transition}

\author{Kosuke Nomura}
\email{nomura@sci.hokudai.ac.jp}
\affiliation{Department of Physics, 
Hokkaido University, Sapporo 060-0810, Japan}
\affiliation{Nuclear Reaction Data Center, 
Hokkaido University, Sapporo 060-0810, Japan}

\date{\today}

\begin{abstract}
The low-energy structure and $\beta$ decay properties 
of the neutron-rich even-mass nuclei near the 
neutron number $N=60$ that are experimentally 
of much interest are investigated within 
the framework of the nuclear density 
functional theory and the 
interacting boson-fermion-fermion model. 
By using the 
results of the constrained self-consistent 
mean-field calculations based on the relativistic 
energy density functional, 
the interacting-boson Hamiltonian 
describing the even-even core 
nuclei, the boson-fermion, and the fermion-fermion 
interactions are determined. 
The Gamow-Teller transition 
strengths, with the corresponding 
operator being constructed 
without introducing further 
phenomenological adjustment, 
are computed by using the 
wave functions of the initial and final 
nuclei of the $\beta$ decay. 
The triaxial quadrupole potential 
energy surfaces computed for the 
$N=60$ even-even isotones suggest a pronounced 
$\gamma$ softness. The calculated energy 
spectra for the even-even and odd-odd nuclei 
in the Rb to Cd isotopic chains 
exhibit an abrupt change in nuclear structure 
around $N=60$, as suggested experimentally. 
The predicted $\beta$-decay $\ft$ values 
underestimate the measured values 
for the nuclei with low $Z$ and with 
$N \leqslant 60$, exhibit a rapid increase 
for $N>60$, reflecting the nuclear structure 
evolution, and agree rather 
well with the measured values 
for those nuclei with $Z$ being 
not far from the proton major shell 
closure $Z=50$. 
Sensitivity of the predicted 
$\beta$-decay properties to the model assumptions 
and parameters employed in the nuclear structure 
calculations is discussed, specifically, 
by comparing results 
obtained based on the different choices of 
the underlying energy density functional. 
\end{abstract}

\maketitle

\section{Introduction}

The low-energy structure of neutron-rich heavy nuclei 
with the neutron number $N\approx$ 60 and 
with the mass $A \approx 100$ has been of much interest 
from both experimental and theoretical points of 
view. In these nuclear systems 
a subtle interplay between the 
single-particle and collective degrees of freedom 
plays an essential role. 
The nuclear structure phenomena 
that are extensively studied 
in this mass region include the abrupt change of 
nuclear shapes around $N=60$, often referred to 
as quantum phase transitions \cite{cejnar2010}, 
the coexistence of different intrinsic 
shapes \cite{heyde2011} 
in the neighborhood of the ground state and 
corresponding low-lying excited $0^+$ states. 
Experiments using radioactive-ion beams have been 
carried out to reveal properties of 
those neutron-rich nuclei 
that are even heavier and beyond the region of 
shape phase transitions at $N\approx$ 60. 
Theoretical predictions for the 
neutron-rich nuclei in this region 
have been made extensively with 
a number of nuclear structure models, 
such as the nuclear shell model 
\cite{caurier2005,sieja2009,togashi2016,shimizu2017}, 
methods based on the nuclear 
energy density functional (EDF) 
\cite{bender2003,vretenar2005,niksic2011,mei2012,robledo2019}, 
and the interacting boson model (IBM) 
\cite{IBM,nomura2016zr,garciaramos2020,nomura2020zr,gavrielov2022}.

Along with their nuclear structure aspects, 
the neutron-rich heavy nuclei 
are also relevant to astrophysical nucleosynthesis 
processes, i.e., the rapid 
neutron-capture ($r$) process and $\beta$ decay. 
The $\beta$-decay rate should 
be sensitive to the wave functions for the 
initial and final nuclei of the process, 
which vary significantly 
from one nucleus to another 
in the transitional regions. 
Reliable theoretical descriptions, 
as well as precise measurements 
\cite{dillmann2003,nishimura2011,quinn2012,lorusso2015,caballero2016}, 
are key to evaluate the nuclear 
$\beta$-decay matrix elements 
and hence to model the astrophysical 
process producing heavy chemical elements. 
Consistent descriptions 
of the nuclear low-lying structure and $\beta$ decay 
have been provided by several theoretical approaches 
such as the interacting boson-fermion and 
boson-fermion-fermion models (IBFM and IBFFM) 
\cite{navratil1988,DELLAGIACOMA1989,brant2004,yoshida2013,mardones2016,nomura2020beta-1,nomura2020beta-2,ferretti2020}, 
the quasiparticle random-phase approximations 
\cite{alvarez2004,sarriguren2015,boillos2015,pirinen2015,simkovic2013,mustonen2016,suhonen2017,ravlic2021,yoshida2023}, 
and the nuclear shell model 
\cite{langanke2003,caurier2005,syoshida2018,suzuki2018}.

Furthermore, what is worth 
mentioning is the double-$\beta$ decay, 
a rare process in which single-$\beta$ decay 
occurs successively 
between the neighboring even-even nuclei, 
emitting two electrons (or positrons) and 
some light particles such as neutrinos 
\cite{avignone2008,engel2017,agostini2023}. 
In particular, should that type of the 
double-$\beta$ decay that does not emit 
neutrinos 
(i.e., neutrinoless double-$\beta$ decay; $\znbb$) 
be observed by experiment, it would 
provide crucial pieces of information 
about the masses and the nature of neutrinos, 
and the validity of various symmetry 
requirements for the electroweak 
fundamental interaction. 
Since the predicted $\znbb$ 
nuclear matrix elements (NMEs) differ 
by several factors among different 
theoretical approaches, 
tremendous efforts have been devoted 
to reduce and control the theoretical 
uncertainties inherent to the models employed. 
The study of the single-$\beta$ decay 
is also instrumental for predicting accurately 
the double-$\beta$ NMEs, especially when 
it is necessary to compute intermediate 
states of the odd-odd nuclei without 
assuming the closure approximation.

Among the theoretical models describing 
$\beta$ decay, as well as nuclear structure, 
the IBM, a model in which correlated monopole and 
quadrupole pairs of valence nucleons are represented 
by $s$ and $d$ bosons, respectively \cite{IBM,OAI,OAIT}, 
has been quite successful in the quantitative 
description of the quadrupole collective states 
of medium-heavy and heavy even-even nuclei. 
In the conventional IBM studies, calculations 
have been purely phenomenological, since 
the parameters of the Hamiltonian have been 
obtained from experiment. 
On the other hand, the IBM should be 
rooted in the underlying microscopic 
nuclear structure, and the model Hamiltonian 
has been shown to be derived from 
nucleonic degrees of freedom 
\cite{IBM,OAIT,OAI,mizusaki1997,nomura2008}. 
In particular, a fermion-to-boson 
mapping technique has been developed 
\cite{nomura2008}, 
in which the IBM Hamiltonian is completely 
determined by using the results of 
the self-consistent mean-field (SCMF) 
calculation that is based on a given EDF 
This method has been applied to a number 
of studies on the quadrupole collective states 
\cite{nomura2008,nomura2010,nomura2011rot,nomura2012tri}. 
An extension of the method to those nuclear 
systems with odd numbers of neutrons and/or protons 
has been made by incorporating 
the particle-boson coupling, 
with microscopic input provided 
by the same EDF calculations 
\cite{nomura2016odd,nomura2019dodd}. 
In these cases, in addition 
to the IBM Hamiltonian describing an 
even-even nucleus, unpaired nucleon degrees 
of freedom and their coupling to the even-even 
boson core should be considered in the 
framework of the IBFM or IBFFM. 
The IBFM and IBFFM formulated in that way 
have also been employed to study $\beta$ decay 
of the $\gamma$-soft nuclei near the Ba and Xe 
regions \cite{nomura2020beta-1,nomura2020beta-2}, 
the neutron-deficient Ge and As nuclei 
\cite{nomura2022beta-ge}, 
the neutron-rich Pd and Rh nuclei 
\cite{nomura2022beta-rh}, and the 
two-neutrino double-$\beta$ ($\tnbb$) 
decay in a large number of 
the candidate nuclei \cite{nomura2022bb}.

This article presents a simultaneous description 
of the low-energy collective excitations 
and $\beta$-decay properties of the 
neutron-rich nuclei in the vicinity
of $N=60$, 
which is experimentally of much interest, 
within the mapped IBM framework mentioned above. 
The scope of the analysis 
is to study the correlations between the 
changes in the nuclear structure of 
the initial and final even-even nuclei 
and the predictions of the $\beta$ decay. 
The present study covers the even-even 
and odd-odd nuclei 
from the proton number $Z$ = 36 (Kr) 
to 48 (Cd) isotopes with the neutron 
number $54 \leqslant N \leqslant 64$, 
in which region rapid shape phase 
transitions are expected to occur. 
It is noted that the 
present analysis is restricted to the 
allowed $\beta$ decay, i.e., the transition in which 
parity is conserved and that takes place 
between those states with angular momenta $I$
differing by $\Delta I \leqslant 1$. 
Note also that only the $\beta$ decay between 
positive-parity states is considered. 
In addition, some of the nuclei included 
in the present analysis are also candidates 
for the $\znbb$ decay, e.g., $^{96}$Zr 
and $^{100}$Mo. Their structure and single-$\beta$ 
and double-$\beta$ properties have already been 
studied in a previous paper 
\cite{nomura2022bb}, and some updated 
results on these particular nuclei 
are included in the present article.

The paper is organized as follows. 
In Sec.~\ref{sec:theory}, the theoretical 
framework to describe low-lying states of the 
considered even-even and odd-odd nuclei 
is presented, followed by 
the definition of the $\beta$-decay 
operator. In Sec.~\ref{sec:results}, 
results of the SCMF calculations along the $N=60$ 
isotones, and of the spectroscopic 
calculations on the low-spin and low-energy spectra, 
and some electromagnetic transition properties 
of the considered nuclei 
are discussed. 
In Sec.~\ref{sec:beta}, results of the 
$\beta$-decay properties, and the sensitivity 
of the final results to the choice of the 
EDF, are discussed. 
A summary of the main results and conclusions 
are given in Sec.~\ref{sec:summary}.

\section{Theoretical framework\label{sec:theory}}

\subsection{Self-consistent mean-field method}

As the first step, the constrained SCMF calculations 
are performed for a set of even-even Kr, Sr, Zr, Mo, 
Ru, Pd, and Cd isotopes with 
$54\leqslant N \leqslant 64$ by means of the  
relativistic Hartree-Bogoliubov (RHB) method 
\cite{vretenar2005,niksic2011,DIRHB,DIRHBspeedup}. 
For the particle-hole channel, 
the density-dependent point coupling (DD-PC1) 
EDF \cite{DDPC1} is employed, which is widely used 
in microscopic calculations on nuclear structure 
and dynamics \cite{niksic2011}. 
For the particle-particle channel the separable pairing 
force of finite range \cite{tian2009} 
is considered with the strength 728 MeV fm$^3$, 
which has been determined to reproduce 
the pairing gap obtained from the 
Gogny D1S \cite{D1S} interaction. 
The sensitivity of the mean-field results on 
various intrinsic properties 
to the choice of relativistic EDF and to the 
strength of this pairing force 
has been extensively studied, e.g., 
in Ref.~\cite{teeti2021}.

Here the particular choice of the 
relativistic EDF and the pairing interaction 
is made as the microscopic input to the mapped IBM-2, 
since these combinations have been shown to be 
adequate in a number 
of nuclear structure studies in different 
mass regions concerning, e.g., 
the octupole deformations and collectivity 
\cite{nomura2013oct,nomura2014,nomura2018oct}, 
and the structural evolution in 
neutron-rich even-even and odd-mass Zr 
isotopes \cite{nomura2020zr}, and in the 
systematic calculation of the 
double-$\beta$ decay of 26 even-even nuclei 
\cite{nomura2022bb} 
and single-$\beta$ decay in transitional 
As and Ge nuclei \cite{nomura2022beta-ge}. 
As in the case of these previous RHB plus IBM-2 studies, 
the constraints imposed in the present 
RHB SCMF calculations 
are on the expectation values of the mass 
quadrupole moments $\hat Q_{20}$ and $\hat Q_{22}$, 
which are related to the polar deformation 
variables $\beta$ and $\gamma$ representing 
degrees of axial deformation and triaxiality, 
respectively \cite{BM}. 
The SCMF calculations provide 
intrinsic properties such as 
the potential energy surfaces (PESs) 
defined in terms of the 
triaxial quadrupole deformations $(\beta,\gamma)$, 
and single-particle energies and occupation 
probabilities. 
These quantities are used 
as a microscopic input for the 
spectroscopic calculations, as described below.

\subsection{Interacting boson-fermion-fermion model}

Within the mean-field approximations 
some important symmetries, such as the 
rotational invariance and conservation 
of particle numbers, are broken. 
To study physical observables in the 
laboratory frame such as the excitation energies 
and electromagnetic transition rates, 
it is necessary to go beyond 
the SCMF level \cite{RS}, by taking into 
account the dynamical 
correlations arising from the restorations 
of symmetries and from the inclusion of quantum 
fluctuations around the mean-field solution, 
which are not taken into account properly 
in the mean-field approximation. 
Such a beyond-mean-field treatment  
is here made by means of the IBM.

In the present analysis 
the neutron-proton version 
of the IBM (IBM-2) is considered, 
which consists of the neutron and proton 
bosons reflecting the collective neutron and proton 
pairs from a microscopic point of view \cite{IBM,OAI,OAIT}. 
The numbers of neutron, $N_{\nu}$, 
and proton, $N_{\pi}$, bosons 
are conserved separately, and are equal to 
half the numbers of valence neutron and proton 
pairs, respectively. 
Here the $^{78}$Ni doubly magic nucleus 
is taken as the inert core for describing 
the even-even Kr, Sr, Zr, Mo, Ru, Pd, and Cd 
nuclei. 
The distinction between the neutron and proton 
bosons is made also in the IBFFM, 
denoted hereafter as IBFFM-2. 
The IBFFM-2 Hamiltonian is given in general as
\begin{align}
\label{eq:ham}
 \hat{H}=\hb + \hf^{\nu} + \hf^{\pi} + \hbf^{\nu} + \hbf^{\pi} + \hff \; .
\end{align}
The first term on the right-hand side 
of the above equation 
denotes the IBM-2 Hamiltonian describing the 
even-even nucleus, and is of the form
\begin{align}
\label{eq:bham}
 \hb = 
&\epsilon_{d}(\hat{n}_{d_{\nu}}+\hat{n}_{d_{\pi}})
+\kappa\hat{Q}_\nu \cdot\hat{Q}_\pi
\nonumber\\
&+\kappa_{\nu}\hat{Q}_\nu \cdot\hat{Q}_\nu
+\kappa_{\pi}\hat{Q}_\pi \cdot\hat{Q}_\pi
+ \kappa'\hat{L}\cdot\hat{L} \; ,
\end{align}
where the first term 
stands for the $d$-boson number operator with 
$\hat n_{d_\rho} = d^\+_\rho\cdot\tilde d_\rho$ 
($\rho=\nu$ or $\pi$) and with $\epsilon_d$ 
the single $d$ boson energy.  
The second, third, and fourth terms are 
the quadrupole-quadrupole 
interactions between neutron and proton bosons, 
between neutron and neutron bosons, 
and between proton and proton bosons, 
respectively. 
The quadrupole operator $\hat Q_{\rho}$ is 
defined as 
$\hat Q_{\rho} = s^\+_\rho \tilde d_\rho + d^\+ s_{\rho} + \chi_{\rho} (d^\+_\rho\times\tilde d_\rho)^{(2)}$, 
with $\chi_\nu$ and $\chi_\pi$ being
dimensionless parameters. 
$\kappa$, $\kappa_{\nu}$, and $\kappa_\pi$ 
are strength parameters.

Among the quadrupole-quadrupole interactions, 
the unlike-boson interaction, 
$\hat{Q}_\nu \cdot\hat{Q}_\pi$, makes 
a dominant contribution to 
low-lying collective states. 
For most of the nuclei considered in this study, 
the like-boson quadrupole-quadrupole interaction terms, 
$\hat{Q}_\nu \cdot\hat{Q}_\nu$ and 
$\hat{Q}_\pi \cdot\hat{Q}_\pi$, 
turn out to play a minor role, 
and are thus neglected by setting 
$\kappa_\nu = \kappa_\pi = 0$. 
These terms are, however, included 
for all the Zr isotopes considered and 
$^{100}$Sr and $^{102}$Sr isotopes. 
This choice is based on the peculiar nuclear 
structure and its evolution as a function of $N$ 
in the neutron-rich Zr isotopes, 
as investigated in the earlier mapped IBM-2 with 
RHB calculations reported in Ref.~\cite{nomura2020zr}. 
In that study, 
these like-boson quadrupole-quadrupole 
interaction terms, 
as well as the unlike-boson one, 
have been introduced to better describe 
the low-energy structure at the quantitative detail. 
Since the RHB PESs for the $^{100}$Sr 
and $^{102}$Sr nuclei 
are more or less similar in topology 
to those for the neighboring Zr isotopes, 
the same interaction terms are introduced. 
In the following, for the above-mentioned nuclei 
a simple relation, 
$\kappa_\nu = \kappa_\pi = \kappa/2$, 
is assumed in order to reduce the 
number of parameters.

The fifth term on the right-hand side 
of Eq.~(\ref{eq:bham}) 
stands for a rotational term, 
with $\kappa'$ being 
the strength parameter, and 
$\hat L = \hat L_{\nu} + \hat L_{\pi}$
denotes the angular momentum operator 
with $\hat L_\rho = (d^\+_\rho\times\tilde d^\+_\rho)^{(1)}$.

The second and third terms of 
Eq.~(\ref{eq:ham}) represent 
the single-neutron and -proton Hamiltonians, 
respectively, and take the form
\begin{align}
\label{eq:hf}
 \hf^{\rho} = -\sum_{\jr}\epsilon_{\jr}\sqrt{2\jr+1}
  (a_{\jr}^\+\times\tilde a_{\jr})^{(0)}
\equiv
\sum_{\jr}\epsilon_{\jr}\hat{n}_{\jr} \; ,
\end{align}
where $\epsilon_{\jr}$ stands for the 
single-particle energy of the odd neutron $(\rho=\nu)$ 
or proton ($\rho=\pi$) orbital $\jr$. 
$a_{\jr}^{(\+)}$ represents 
the particle annihilation (creation) operator, 
with $\tilde{a}_{\jr}$ defined by 
$\tilde{a}_{\jr\mr}=(-1)^{\jr -\mr}a_{\jr-\mr}$. 
On the right-hand side of Eq.~(\ref{eq:hf}), 
$\hat{n}_{\jr}$ stands for the number operator 
for the odd particle. 
The single-particle space taken in the 
present study comprises the neutron $3s_{1/2}$, 
$2d_{3/2}$, $2d_{5/2}$, and $1g_{7/2}$ orbitals, 
and the proton $1g_{9/2}$ orbital in the 
$N=50-82$ and $Z=28-50$ 
major oscillator shells for calculating 
the positive-parity states of the 
odd-odd nuclei.

The fourth (fifth) term on the 
right-hand side  
of Eq.~(\ref{eq:ham}) denotes the interaction 
between a single neutron (or proton) and 
the even-even boson core. 
A simplified form which was derived 
microscopically within the generalized 
seniority scheme \cite{IBFM,scholten1985} 
is here adopted: 
\begin{equation}
\label{eq:hbf}
 \hbf^{\rho}
=\Gamma_{\rho}\hat{V}_{\mathrm{dyn}}^{\rho}
+\Lambda_{\rho}\hat{V}_{\mathrm{exc}}^{\rho}
+A_{\rho}\hat{V}_{\mathrm{mon}}^{\rho} \; ,
\end{equation}
where the first, second, and third terms 
represent the quadrupole dynamical, exchange, 
and monopole interactions, respectively, 
with the strength parameters 
$\Gamma_\rho$, $\Lambda_\rho$, and $A_{\rho}$. 
Each term in the above expression reads
\begin{widetext}
\begin{align}
\label{eq:dyn}
&\hat{V}_{\mathrm{dyn}}^{\rho}
=\sum_{\jr\jr'}\gamma_{\jr\jr'}
(a^{\+}_{\jr}\times\tilde{a}_{\jr'})^{(2)}
\cdot\hat{Q}_{\rho'},\\
\label{eq:exc}
&\hat{V}^{\rho}_{\mathrm{exc}}
=-\left(
s_{\rho'}^\+\times\tilde{d}_{\rho'}
\right)^{(2)}
\cdot
\sum_{\jr\jr'\jr''}
\sqrt{\frac{10}{N_{\rho}(2\jr+1)}}
\beta_{\jr\jr'}\beta_{\jr''\jr}
:\left[
(d_{\rho}^{\+}\times\tilde{a}_{\jr''})^{(\jr)}\times
(a_{\jr'}^{\+}\times\tilde{s}_{\rho})^{(\jr')}
\right]^{(2)}:
+ (\text{H.c.}) \; ,\\
\label{eq:mon}
&\hat{V}_{\mathrm{mon}}^{\rho}
=\hat{n}_{d_{\rho}}\hat{n}_{\jr} \; ,
\end{align}
\end{widetext}
where the $j$-dependent factors 
$\gamma_{\jr\jr'}=(u_{\jr}u_{\jr'}-v_{\jr}v_{\jr'})Q_{\jr\jr'}$, 
and $\beta_{\jr\jr'}=(u_{\jr}v_{\jr'}+v_{\jr}u_{\jr'})Q_{\jr\jr'}$, 
with 
$Q_{\jr\jr'}=\braket{\ell_{\rho}\frac{1}{2}\jr\|Y^{(2)}\|\ell'_\rho\frac{1}{2}\jr'}$ being the matrix element of the fermion 
quadrupole operator in the single-particle basis. 
$\hat{Q}_{\rho'}$ in Eq.~(\ref{eq:dyn}) denotes 
the quadrupole operator in the boson system, 
and was already introduced in Eq.~(\ref{eq:bham}). 
The notation $:(\cdots):$ in Eq.~(\ref{eq:exc}) 
stands for normal ordering. 
Note that the forms of $\hbf^{\nu}$ 
and $\hbf^\pi$ are considered based on the 
assumption \cite{scholten1985,IBFM} 
that among the boson-fermion 
interactions those between unlike particles 
[i.e., between a neutron (proton) and 
proton (neutron) bosons] are most important 
for the quadrupole dynamical and exchange 
terms, and those between like particles 
[i.e., between a neutron (proton) and 
neutron (proton) bosons] are relevant for 
the monopole term. 
It is also noted that within the seniority 
considerations the unperturbed single-particle 
energy for the orbital $\jr$, 
$\epsilon_{\jr}$, in Eq.~(\ref{eq:hf}) 
should be replaced with the quasiparticle 
energy, denoted $\tilde\epsilon_{\jr}$.

The last term of the IBFFM-2 
Hamiltonian (\ref{eq:ham}), $\hff$, 
corresponds to the residual interaction 
between the unpaired neutron and proton. 
The following form is considered 
for this interaction. 
\begin{align}
\label{eq:hff}
\hff
=& 4\pi({\vd} + \vssd {\bm{\sigma}}_{\nu}
\cdot{\bm{\sigma}}_{\pi} )
\delta(\bm{r})
\delta(\bm{r}_{\nu}-r_0)
\delta(\bm{r}_{\pi}-r_0)
\nonumber\\
& - \frac{1}{\sqrt{3}} \vsss {\bm{\sigma}}_{\nu}
\cdot{\bm{\sigma}}_{\pi}
+ \vt
\left[
\frac{3({\bm\sigma}_{\nu}\cdot{\bf r})
({\bm\sigma}_{\pi}\cdot{\bf r})}{r^2}
-{\bm{\sigma}}_{\nu}
\cdot{\bm{\sigma}}_{\pi}
\right] \; .
\end{align}
The first term consists of the $\delta$, and 
spin-spin $\delta$ terms, while the second 
and third terms represent, respectively, 
the spin-spin 
and tensor interactions. 
$\vd$, $\vssd$, $\vsss$, and $\vt$
are strength parameters. 
Note that $\bm{r}=\bm{r}_{\nu}-\bm{r}_{\pi}$ 
is the relative coordinate of the 
neutron and proton, and $r_0=1.2A^{1/3}$ fm. 
The matrix element of $\hff$, denoted by 
$V_{\nu\pi}'$, has the following 
$(u,v)$-dependent form
\cite{yoshida2013}: 
\begin{align}
\label{eq:vres}
\hff'
&= (u_{j_\nu'} u_{j_\pi'} u_{j_\nu} u_{j_\nu} + v_{j_\nu'} v_{j_\pi'} v_{j_\nu} v_{j_\nu})
V^{J}_{j_\nu' j_\pi' j_\nu j_\pi}
\nonumber \\
& {} - (u_{j_\nu'}v_{j_\pi'}u_{j_\nu}v_{j_\pi} +
 v_{j_\nu'}u_{j_\pi'}v_{j_\nu}u_{j_\pi}) \nonumber \\
&\times \sum_{J'} (2J'+1)
\left\{ \begin{array}{ccc} {j_\nu'} & {j_\pi} & J' \\ {j_\nu} & {j_\pi'} & J
\end{array} \right\} 
V^{J'}_{j_\nu'j_\pi j_\nu j_\pi'} \; ,
\end{align}
with 
\begin{align}
V^{J}_{j_\nu'j_\pi'j_\nu j_\pi} 
= \langle j_\nu'j_\pi';J| \hff |j_\nu j_\pi;J\rangle
\end{align}
being the matrix element between 
the bases defined in terms of 
the neutron-proton pair 
coupled to the angular momentum $J$. 
The bracket in Eq.~(\ref{eq:vres}) 
represents the Racah coefficient. 
By following the procedure of 
Ref.~\cite{morrison1981}, 
those terms resulting from contractions are 
neglected in Eq.~(\ref{eq:vres}).

\subsection{Procedure to build the IBFFM-2 Hamiltonian}

The procedure to determine the IBFFM-2 Hamiltonian 
(\ref{eq:ham}) consists in the following three steps. 
\begin{itemize}
 \item[(i)] 
First, the IBM-2 Hamiltonian 
is determined in such a way that the 
$(\beta,\gamma)$ PES obtained 
from the constrained SCMF calculation 
is mapped onto the corresponding 
one in the IBM system which 
is represented as the energy expectation value 
in the boson coherent state \cite{ginocchio1980}. 
This procedure specifies 
optimal parameters of the IBM-2 
that renders the IBM-2 PES 
as similar as possible to the SCMF one. 
Only the strength parameter $\kappa'$ for the 
$\hat L \cdot \hat L$ term [see Eq.~(\ref{eq:bham})] 
is determined separately 
so that the cranking moment of inertia 
calculated in the boson intrinsic state 
at the equilibrium minimum 
is equal to the Inglis-Belyaev (IB)
moment of inertia obtained with the 
RHB calculation. 
The IB moment of inertia 
is here increased by 30 \%, taking into account 
the fact that it considerably 
underestimates the observed moments of inertia. 
See Refs.~\cite{nomura2008,nomura2010,nomura2011rot} 
for details about the determination 
of the IBM-2 Hamiltonian.

\item[(ii)]
The single-fermion Hamiltonian, 
$\hf$ [Eq.~(\ref{eq:hf})], and boson-fermion 
interactions, $\hbf$ [Eq.~(\ref{eq:hbf})], 
are constructed by using the 
procedure of Ref.~\cite{nomura2016odd}: 
The SCMF RHB calculations are performed 
for the neighboring odd-$N$ or odd-$Z$ 
nucleus with the constraint
on zero deformation to provide  
quasiparticle energies, $\tilde\epsilon_{j_\rho}$, and 
occupation probabilities, $v^2_{j_{\rho}}$, 
at the spherical configuration 
for the odd nucleon at orbitals $j_{\rho}$. 
These quantities are then input to $\hf^\rho$ 
and $\hbf^\rho$, respectively. 
The remaining three coupling constants, 
$\Gamma_\rho$, $\Lambda_\rho$, and $A_\rho$, 
are determined to fit the experimental data 
for a few low-lying positive-parity 
levels of each odd-$N$ and odd-$Z$ nuclei. 
Table~\ref{tab:comp} summarizes the even-even, 
neighboring odd-$N$, odd-$Z$, 
and odd-odd nuclei studied in this paper. 
It is worth mentioning that 
Y nuclei, with $Z=39$, correspond 
to the middle of the proton major shell $Z=28-50$, 
and their even-even boson cores are 
here considered to be Sr nuclei ($Z=38$). 
Alternatively, one may also consider 
Zr cores for Y.

\item[(iii)]
The parameters
$\Gamma_\rho$, $\Lambda_\rho$, and $A_\rho$, 
which are determined by fitting to the neighboring 
odd-$N$ and odd-$Z$ nuclei, are used for the 
odd-odd nucleus. The quasiparticle energies, 
$\tilde\epsilon_{\jr}$,
and occupation probabilities, $v_{\jr}^2$,
are newly calculated. 
Then the interaction strengths 
in the fermion-fermion interaction 
(\ref{eq:hff}) are fixed to reproduce, 
to a certain accuracy, 
the observed low-lying positive-parity 
levels of each odd-odd nucleus. 

\end{itemize}
The IBFFM-2 Hamiltonian, with the parameters 
determined by the above procedure, is diagonalized 
to yield excitation energies and wave functions 
of the odd-odd systems.

\begin{table}
\caption{\label{tab:comp}
The even-even core, neighboring odd-$N$, odd-$Z$, 
and odd-odd nuclei with the neutron number 
$N=54-64$ considered in the present study. 
}
 \begin{center}
 \begin{ruledtabular}
  \begin{tabular}{cccc}
Even-even & Odd-$N$  & Odd-$Z$ & Odd-odd  \\
\hline
$^{}_{48}$Cd$_{N}$ & $^{}_{48}$Cd$_{N+1}$ & $^{}_{47}$Ag$_{N}$ & $^{}_{47}$Ag$_{N+1}$ \\
$^{}_{46}$Pd$_{N}$ & $^{}_{46}$Pd$_{N+1}$ & $^{}_{45}$Rh$_{N}$ & $^{}_{45}$Rh$_{N+1}$ \\
$^{}_{44}$Ru$_{N}$ & $^{}_{44}$Ru$_{N+1}$ & $^{}_{43}$Tc$_{N}$ & $^{}_{43}$Tc$_{N+1}$ \\
$^{}_{42}$Mo$_{N}$ & $^{}_{42}$Mo$_{N+1}$ & $^{}_{41}$Nb$_{N}$ & $^{}_{41}$Nb$_{N+1}$ \\
$^{}_{40}$Zr$_{N}$ & $^{}_{40}$Zr$_{N+1}$ & &  \\
$^{}_{38}$Sr$_{N}$ & $^{}_{38}$Sr$_{N+1}$ & $^{}_{39}$Y$_{N+1}$ & $^{}_{39}$Y$_{N+1}$ \\
$^{}_{36}$Kr$_{N}$ & $^{}_{36}$Kr$_{N+1}$ & $^{}_{37}$Rb$_{N+1}$ & $^{}_{37}$Rb$_{N+1}$ \\
  \end{tabular}
 \end{ruledtabular}
 \end{center}
\end{table}

\subsection{Electromagnetic transition operators}

By using the IBFFM-2 wave functions, 
electromagnetic properties are calculated. 
The $E2$ operator is defined as
\begin{align}
 \label{eq:e2}
\hat T^{(E2)}
= \hat T^{(E2)}_\text{B}
+ \hat T^{(E2)}_\text{F} \; ,
\end{align}
with the boson part, 
\begin{align}
 \label{eq:e2b}
\hat T^{(E2)}_\mathrm{B}
= e_\nu^\mathrm{B}\hat Q_\nu 
+ e_\pi^\mathrm{B}\hat Q_\pi \; ,
\end{align}
and the fermion part, 
\begin{align}
 \label{eq:e2f}
\hat T^{(E2)}_\mathrm{F}
=-\frac{1}{\sqrt{5}}
&\sum_{\rho=\nu,\pi}
\sum_{\jr\jr'}
(u_{\jr}u_{\jr'}-v_{\jr}v_{\jr'})
\nonumber\\
&\times
\left\langle
\ell_\rho\frac{1}{2}\jr 
\bigg\| 
e^\mathrm{F}_\rho r^2 Y^{(2)} 
\bigg\|
\ell_\rho'\frac{1}{2}\jr'
\right\rangle
(a_{\jr}^\dagger\times\tilde a_{\jr'})^{(2)} \; .
\end{align}
$e^\mathrm{B}_\rho$ are the boson effective 
charges, and the common values for the neutrons 
and protons, i.e., 
$e^\mathrm{B}_\nu = e^\mathrm{B}_\pi =0.0577$ $e$b, 
which were used in the previous IBFFM-2 
calculations for the Ge and As nuclei 
\cite{nomura2022beta-ge}, are employed here. 
The neutron and proton effective charges,  
$e^\mathrm{F}_\nu =0.5$ $e$b and
$e^\mathrm{F}_\pi =1.5$ $e$b, 
are exploited also from 
Ref.~\cite{nomura2022beta-ge}. 
The $M1$ transition operator 
$\hat T^{(M1)}$ reads 
\begin{align}
 \label{eq:m1}
\hat T^{(M1)}
=\sqrt{\frac{3}{4\pi}}
&\sum_{\rho=\nu,\pi}
\Biggl[
g_\rho^\mathrm{B}\hat L_\rho
-\frac{1}{\sqrt{3}}
\sum_{\jr\jr'}
(u_{\jr}u_{\jr'}+v_{\jr}v_{\jr'})
\nonumber \\
&\times
\left\langle \jr \| g_l^\rho{\bf \ell}+g_s^\rho{\bf s} 
\| \jr' \right\rangle
(a_{\jr}^\+\times\tilde a_{\jr'})^{(1)}
\Biggr].
\end{align}
The empirical $g$ factors for the neutron and
proton bosons, $g_\nu^\mathrm{B}=0\,\mu_N$ and 
$g_\pi^\mathrm{B}=1.0\,\mu_N$, respectively, are adopted. 
For the neutron (or proton) $g$ factors, the 
free values $g_\ell^\nu=0\,\mu_N$ and $g_s^\nu=-3.82\,\mu_N$
($g_\ell^\pi=1.0\,\mu_N$ and $g_s^\pi=5.58\,\mu_N$) 
are employed, with $g_s^\rho$ quenched by 30 \% 
as is the case of many of the realistic 
IBFFM-2 calculations (e.g., Ref.~\cite{yoshida2013}).

\subsection{Gamow-Teller transition operator}

To study $\beta$-decay properties, 
Gamow-Teller (GT) transition strengths are 
computed. 
Here, the Fermi transitions do not enter 
the $\beta$ decay of the studied nuclei, since 
the single-particle space considered for the 
odd-odd nuclei consists of the neutron 
$3s_{1/2}$, $2d_{3/2}$, $2d_{5/2}$, and 
$1g_{7/2}$ orbitals, and the proton $1g_{9/2}$ 
orbital, and there is no coupling between 
these single-particle orbitals to 
the angular momentum zero that gives rise to 
the Fermi transition.

The GT transition operator is here defined by 
\begin{align}
\label{eq:ogt}
&\hat{T}^{\rm GT}
=\sum_{j_{\nu}j_{\pi}}
\eta_{j_{\nu}j_{\pi}}^{\mathrm{GT}}
\left(\hat P_{j_{\nu}}\times\hat P_{j_{\pi}}\right)^{(1)} \; ,
\end{align}
with the coefficients $\eta$ calculated as
\begin{align}
\label{eq:etagt}
\eta_{j_{\nu}j_{\pi}}^{\mathrm{GT}}
&= - \frac{1}{\sqrt{3}}
\left\langle
\ell_{\nu}\frac{1}{2}j_{\nu}
\bigg\|{\bm\sigma}\bigg\|
\ell_{\pi}\frac{1}{2}j_{\pi}
\right\rangle
\delta_{\ell_{\nu}\ell_{\pi}} \; .
\end{align}
$\hat P_{\jr}$ in Eq.~(\ref{eq:ogt}) 
is a one-particle transfer operator, expressed 
as one of these operators, 
 \begin{align}
\label{eq:creation1}
&A^{\+}_{\jr\mr} = \zeta_{\jr} a_{{\jr}\mr}^{\+}
 + \sum_{\jr'} \zeta_{\jr\jr'} s^{\+}_\rho (\tilde{d}_{\rho}\times a_{\jr'}^{\+})^{(\jr)}_{\mr}
\\
\label{eq:creation2}
&B^{\+}_{\jr\mr}
=\theta_{\jr} s^{\+}_\rho\tilde{a}_{\jr\mr}
 + \sum_{\jr'} \theta_{\jr\jr'} (d^{\+}_{\rho}\times\tilde{a}_{\jr'})^{(\jr)}_{\mr} \; ,
\end{align}
which increases the number of valence nucleons, 
$2N_\rho+n_{\jr}$, by 1, 
and their conjugate operators, 
\begin{align}
\label{eq:annihilation1}
&\tilde{A}_{\jr\mr}
=\zeta_{\jr}^{\ast} \tilde a_{{\jr}\mr}
 + \sum_{\jr'} \zeta_{\jr\jr'}^{\ast} s_\rho ({d}_{\rho}^{\+}\times \tilde a_{\jr'})^{(\jr)}_{\mr} \; , \\
\label{eq:annihilation2}
&\tilde{B}_{\jr\mr}
=-\theta_{\jr}^{\ast} s_\rho {a}^{\+}_{\jr\mr}
 - \sum_{\jr'} \theta_{\jr\jr'}^{\ast} (\tilde d_{\rho}\times{a}_{\jr'}^{\+})^{(\jr)}_{\mr} \; ,
\end{align}
which decreases the valence nucleon number by 1. 
In addition, the operators in Eqs.~(\ref{eq:creation1}) 
and (\ref{eq:creation2}) [Eqs.~(\ref{eq:annihilation1}) 
and (\ref{eq:annihilation2})], respectively, increase 
and decrease the number of like-particle 
(like-hole) nucleons. 
The $\hat{T}^{\rm GT}$ 
operator can be constructed by using two 
of those operators defined in 
Eqs.~(\ref{eq:creation1})--(\ref{eq:annihilation2}), 
depending on the type of the $\beta$ decay 
under study (i.e., $\beta^+$ or $\btm$), and 
on the particle or hole nature of bosons in 
the even-even IBM-2 core. 
To be more specific, for the $\btm$ decay, 
$\hat P_{j_\nu}=\tilde A_{j_{\nu}m_{\nu}}$ and 
$\hat P_{j_\pi}=\tilde B_{j_{\pi}m_{\pi}}$ for 
those nuclei with $Z \geqslant 39$, and 
$\hat P_{j_\nu}=\tilde A_{j_{\nu}m_{\nu}}$ and 
$\hat P_{j_\pi}=\tilde A_{j_{\pi}m_{\pi}}^{\+}$ 
for $Z \leqslant 38$. Their conjugate operators 
correspond to those describing the $\beta^+$ decay or 
electron-capture (EC) process.

The coefficients $\zeta_{j}$, $\zeta_{jj'}$, 
$\theta_{j}$, and $\theta_{jj'}$ 
in Eqs.~(\ref{eq:creation1})--(\ref{eq:annihilation2}) 
are calculated by the following formulas obtained 
within the generalized seniority scheme 
\cite{dellagiacoma1988phdthesis}:
\begin{subequations}
 \begin{align}
\label{eq:zeta1}
\zeta_{\jr}&= 
u_{\jr} \frac{1}{K_{\jr}'}, \\
\label{eq:zeta2}
\zeta_{\jr\jr'}
&= -v_{\jr} 
\beta_{\jr'\jr}
\sqrt{\frac{10}{N_{\rho}(2\jr+1)}}\frac{1}{K K_{\jr}'} , \\ 
\label{eq:theta1}
\theta_{\jr}
&= \frac{v_{\jr}}{\sqrt{N_{\rho}}} 
\frac{1}{K_{\jr}''},\\
\label{eq:theta2}
\theta_{\jr\jr'}
&= u_{\jr} 
\beta_{\jr'\jr}
\sqrt{\frac{10}{2\jr+1}} \frac{1}{K K_{\jr}''} \; .
\end{align}
\end{subequations}
The factors $K$, $K_{\jr}'$, and $K_{\jr}''$ 
read
\begin{subequations} 
\begin{align}
&K = \left( \sum_{\jr\jr'} 
\beta_{\jr\jr'}^{2} \right)^{1/2},\\
&K_{\jr}' = \left[ 1 + 2 
\left(\frac{v_{\jr}}{u_{\jr}}\right)^{2} \frac{\braket{(\hat 
n_{s_\rho}+1)\hat n_{d_\rho}}_{0^+_1}} {N_\rho(2\jr+1)} \frac{\sum_{\jr'} 
\beta_{\jr'\jr}^{2}}{K^{2}} \right]^{1/2} ,\\
&K_{\jr}'' = \left[ 
\frac{\braket{\hat n_{s_\rho}}_{0^+_1}}{N_\rho} 
+2\left(\frac{u_{\jr}}{v_{\jr}}\right)^{2} \frac{\braket{\hat 
n_{d_\rho}}_{0^+_1}}{2\jr+1} \frac{\sum_{\jr'} \beta_{\jr'\jr}^{2}}{K^{2}} 
\right]^{1/2},
\end{align} 
\end{subequations}
where $\hat n_{s_{\rho}}$ is the number operator 
for the $s_\rho$ boson and $\braket{\cdots}_{0^+_1}$ 
represents the expectation 
value in the ground state of the even-even core. 
The occupation, $v_{\jr}$, and unoccupation, 
$u_{\jr}$, amplitudes in the above expressions 
are the same as those used when constructing 
the IBFFM-2 Hamiltonian.

It is noted that the operator, $\hat P_{\jr}$, 
adopted in the present study is of a specific 
form, as it depends only on the amplitudes 
$v_{\jr}$ and $u_{\jr}$. 
One could consider more complicated 
one-particle transfer operators 
that are parametrized in some different ways or 
that include some higher-order terms, 
especially because the final results on the 
$\beta$-decay properties are supposed 
to be influenced by the form of the operator. 
Such a more general form of the transfer operator 
would also involve additional parameters 
that are needed for a better description of the experimental data, 
which, in turn, could cause further complications and 
theoretical uncertainties. 
The use of the simplified operators in 
Eqs.~(\ref{eq:creation1})--(\ref{eq:annihilation2}), 
specified by the $v_{\jr}$ and $u_{\jr}$ 
amplitudes only, has an advantage of not having 
to introduce any phenomenological 
parameter in the calculation of $\beta$ decay.

\begin{figure*}[ht]
\begin{center}
\includegraphics[width=.48\linewidth]{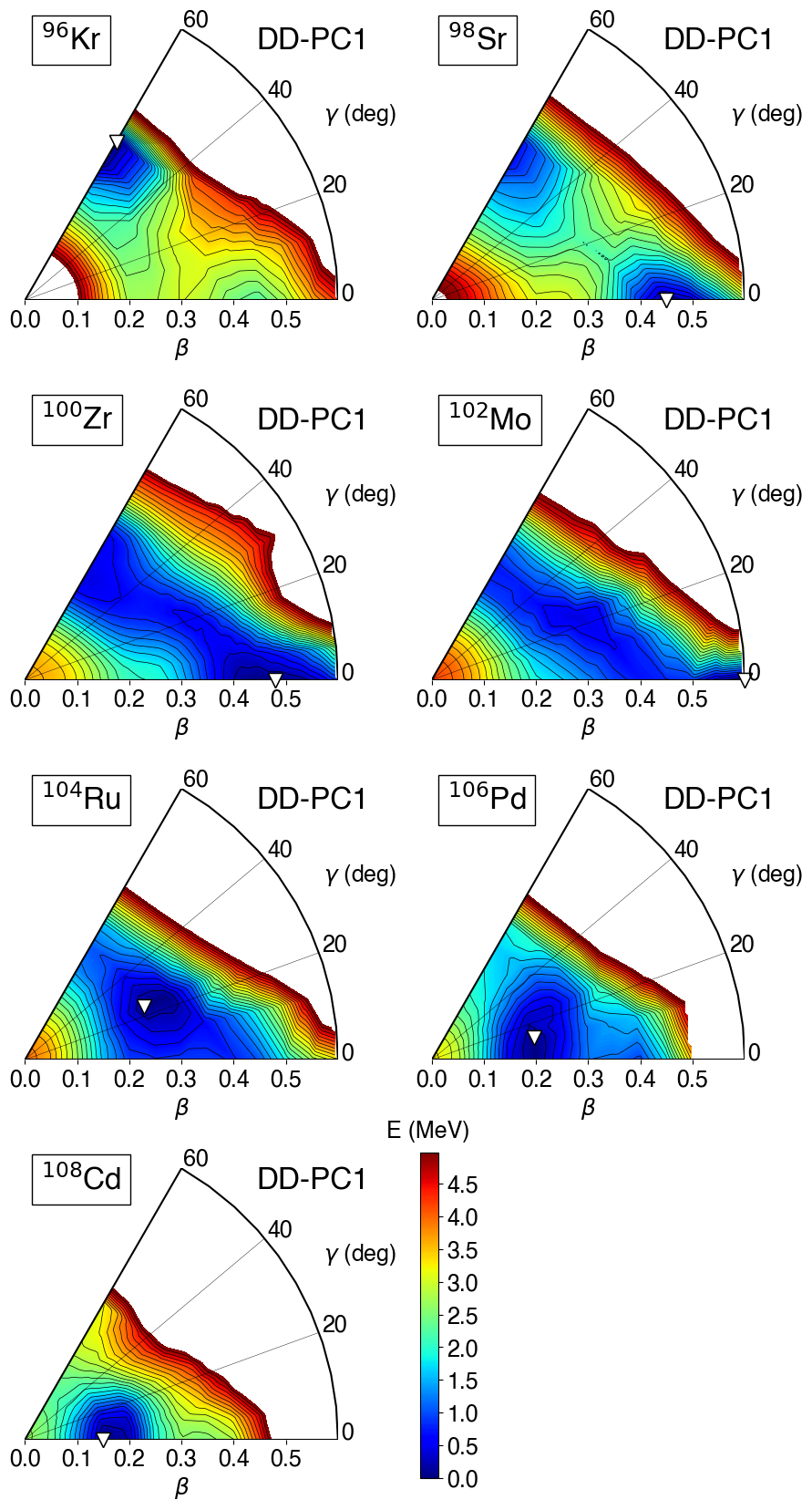}
\includegraphics[width=.48\linewidth]{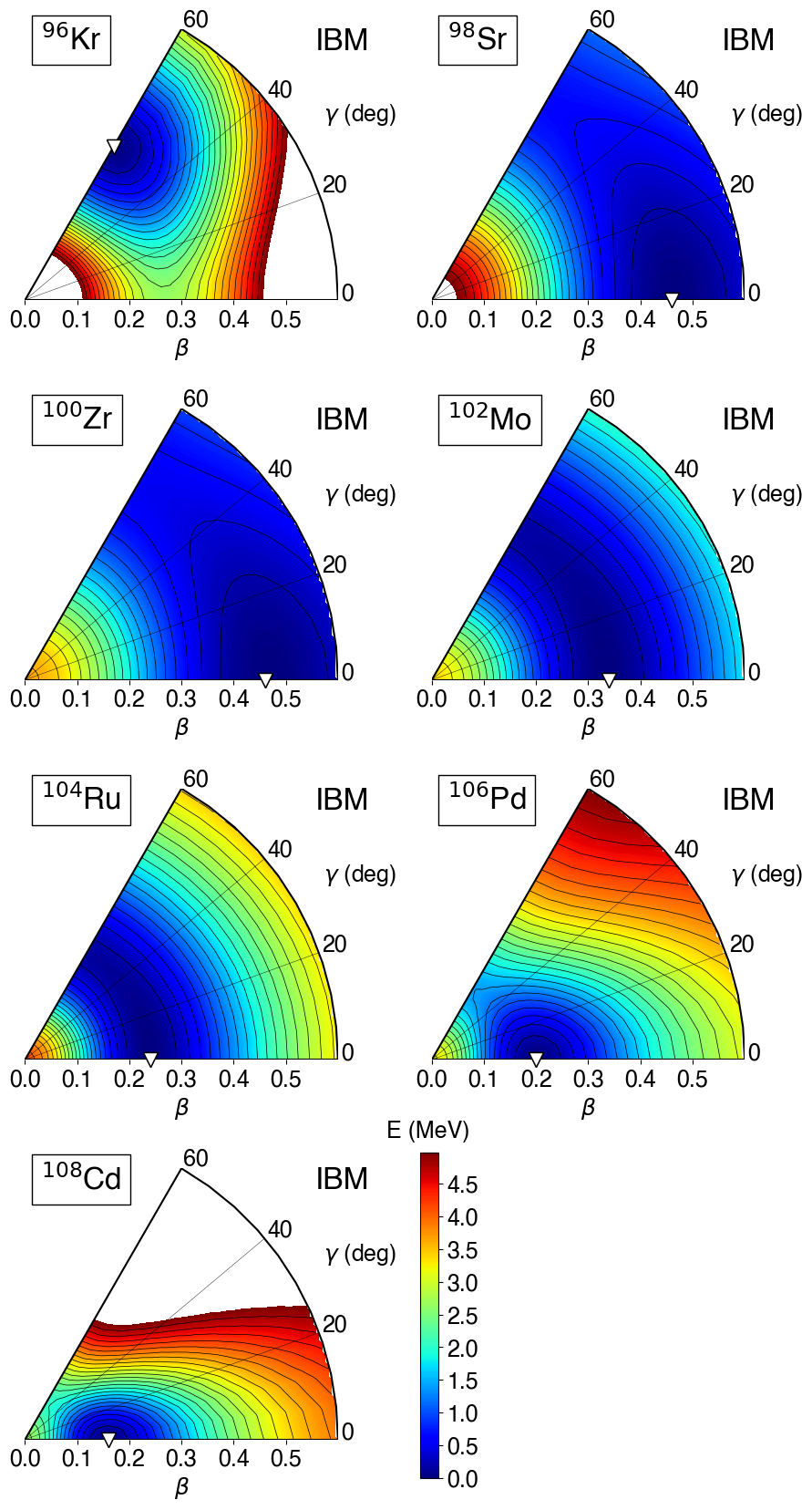}
\caption{Columns 1 and 2: 
Contour plots of the 
triaxial quadrupole $(\beta,\gamma)$ 
PESs for the even-even 
$N=60$ isotones calculated within the constrained 
relativistic Hartree-Bogoliubov method employing 
the energy functional DD-PC1 and 
separable pairing force of finite range. 
Columns 3 and 4: 
The corresponding mapped PESs in the IBM-2. 
The global minimum is 
indicated by an open triangle, and 
the difference between the neighboring 
contours is 0.2 MeV.}
\label{fig:pes}
\end{center}
\end{figure*}

%
\begin{figure*}
\begin{center}
\includegraphics[width=\linewidth]{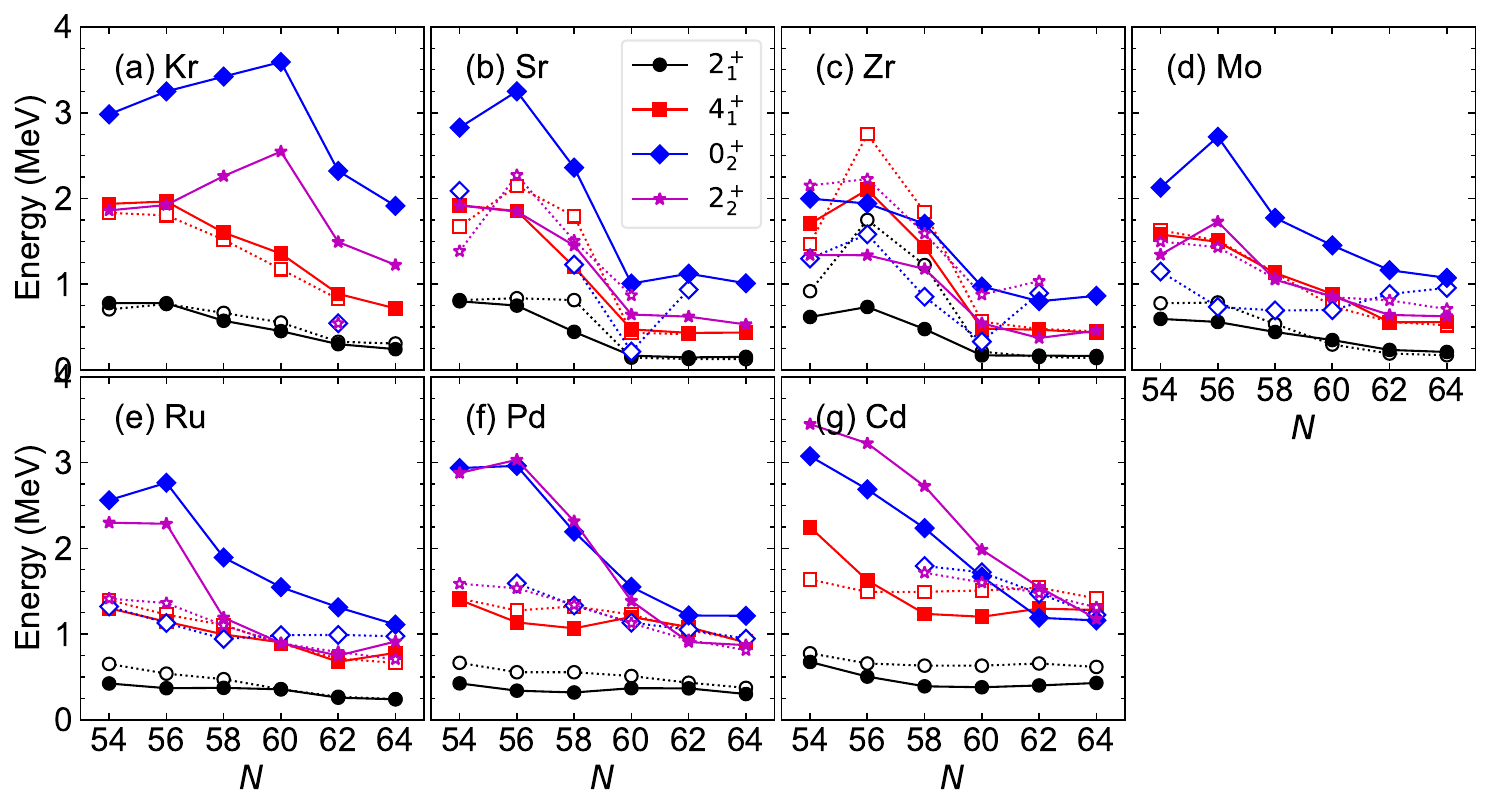}
\caption{Calculated excitation spectra 
of the $2^+_1$, $4^+_1$, 
$0^+_2$, and $2^+_2$ states of the even-even 
Kr, Sr, Zr, Mo, Ru, Pd, and Cd isotopes 
(solid symbols connected by solid lines). 
The experimental data 
(open symbols connected by dotted lines) 
are adopted from the NNDC database \cite{data}.}
\label{fig:ee-level}
\end{center}
\end{figure*}

\section{Low-lying structure of the initial and final nuclei\label{sec:results}}

\subsection{Potential energy surfaces}

In the first and second columns of 
Fig.~\ref{fig:pes} the triaxial quadrupole 
$(\beta,\gamma)$ PESs  
for the $N=60$ isotones, from $^{96}$Kr to $^{108}$Cd, 
computed within the constrained RHB method with 
the DD-PC1 EDF and a separable pairing force, 
are shown. For $^{96}$Kr and $^{98}$Sr the 
SCMF calculation predicts two minima 
on the oblate and prolate sides of the PESs. 
The energy surfaces for 
$^{100}$Zr, $^{102}$Mo, and $^{104}$Ru 
are particularly soft in $\gamma$ deformation. 
For $^{104}$Ru a triaxial minimum around 
$\gamma\approx 30^\circ$ is found. 
The nuclei $^{106}$Pd and $^{108}$Cd 
are suggested to be more weakly (prolate) 
deformed, as expected from the fact that 
they are rather close to 
the proton $Z=50$ major shell closure.

The SCMF PESs can be compared with the 
mapped IBM-2 PESs, which are shown in 
the third and fourth columns of 
Fig.~\ref{fig:pes}. One could observe certain 
similarities between the IBM-2 and SCMF PESs, 
in that basic characteristics of the latter 
in the neighborhood of the global minimum 
are reproduced in the former. 
The difference between the SCMF and IBM-2 
PESs becomes visible 
for those configurations that correspond 
to large $\beta$ deformation, so that the 
IBM-2 surface is flat as compared to the 
SCMF one. This is due to the restricted 
degrees of freedom in the IBM-2 framework; 
that is, the IBM-2 is built only on the 
valence nucleons in one major oscillator shell 
while the SCMF model comprises 
all nucleons.

Another notable difference between 
the SCMF and IBM-2 
PESs is that the former exhibits 
several minima that are close in energy 
to each other, most spectacularly in $^{96}$Kr 
and $^{98}$Sr, whereas a single minimum is 
found in the IBM-2 PES in both cases. 
Here it is assumed that low-lying 
states in the vicinity of the ground state 
are determined mainly by the configurations 
near the absolute minimum of the SCMF PES, and 
hence the mapping is carried out so as to reproduce 
the topology of only that region of the SCMF PES. 
This assumption may not hold for those nuclei 
for which multiple mean-field minima are found. 
Particularly in $^{98}$Sr, 
even though the oblate and prolate mean-field 
minima differ in energy only by 0.67 MeV, 
the mapping is carried out only to reproduce the 
prolate global minimum at $\beta=0.45$. 
The IBM-2 Hamiltonian 
for $^{98}$Sr could be constructed 
based on the oblate secondary minimum at $\beta=0.3$. 
In that case the nature of the 
wave function would be considerably different 
from otherwise, and this would severely affect 
the $\beta$-decay properties. 
It should be also noted that the oblate-prolate 
balance in the energy surface is so subtle 
that it depends on the nuclear EDF that 
underlies the IBM-2 calculation. 
These possibilities should be thoroughly 
investigated in future.

There is, nevertheless, a way of producing the multiple 
minima on the energy surface within the IBM-2, 
that is, by the inclusion of the 
configuration mixing between several different 
boson spaces differing in boson number by two 
\cite{duval1981}. 
Alternatively, cubic, or three-body, 
boson terms with negative strength parameter 
could be introduced 
in the IBM-2 Hamiltonian \cite{nomura2020zr}, 
which are also expected to produce two minima 
on the prolate and oblate sides. 
These extensions are, however, not attempted in the 
present work, mainly because both the configuration mixing 
and the inclusion of the cubic terms cannot 
be handled with the current version of 
the IBFFM-2 code.

Furthermore, as noted earlier, 
the SCMF PES for $^{104}$Ru 
exhibits a triaxial minimum at 
$\gamma\approx 30^\circ$, while the IBM-2 one 
does not. 
This discrepancy could be solved by 
including the three-body boson terms with 
a positive strength parameter \cite{nomura2012tri}. 
By the inclusion of these terms, 
the observed level structure and $E2$ transition properties 
of the $\gamma$-vibrational bands of 
$\gamma$-soft nuclei are shown \cite{nomura2012tri} 
to be better reproduced, but the energy levels 
of the ground-state and excited $0^+$ bands 
are not altered significantly. 
For this reason, and since 
in the following discussion mainly the $\beta$ decay 
that involves the $0^+_1$ ground states 
of the even-even nuclei is considered, 
the three-body boson terms are 
neglected in the present calculation. 
In some cases, however, the triaxiality 
could make non-negligible contributions to 
the nuclear wave functions, 
and it will be of great interest to investigate 
the roles played by the quadrupole triaxiality 
in the $\beta$-decay properties.

%
\begin{figure}
\begin{center}
\includegraphics[width=\linewidth]{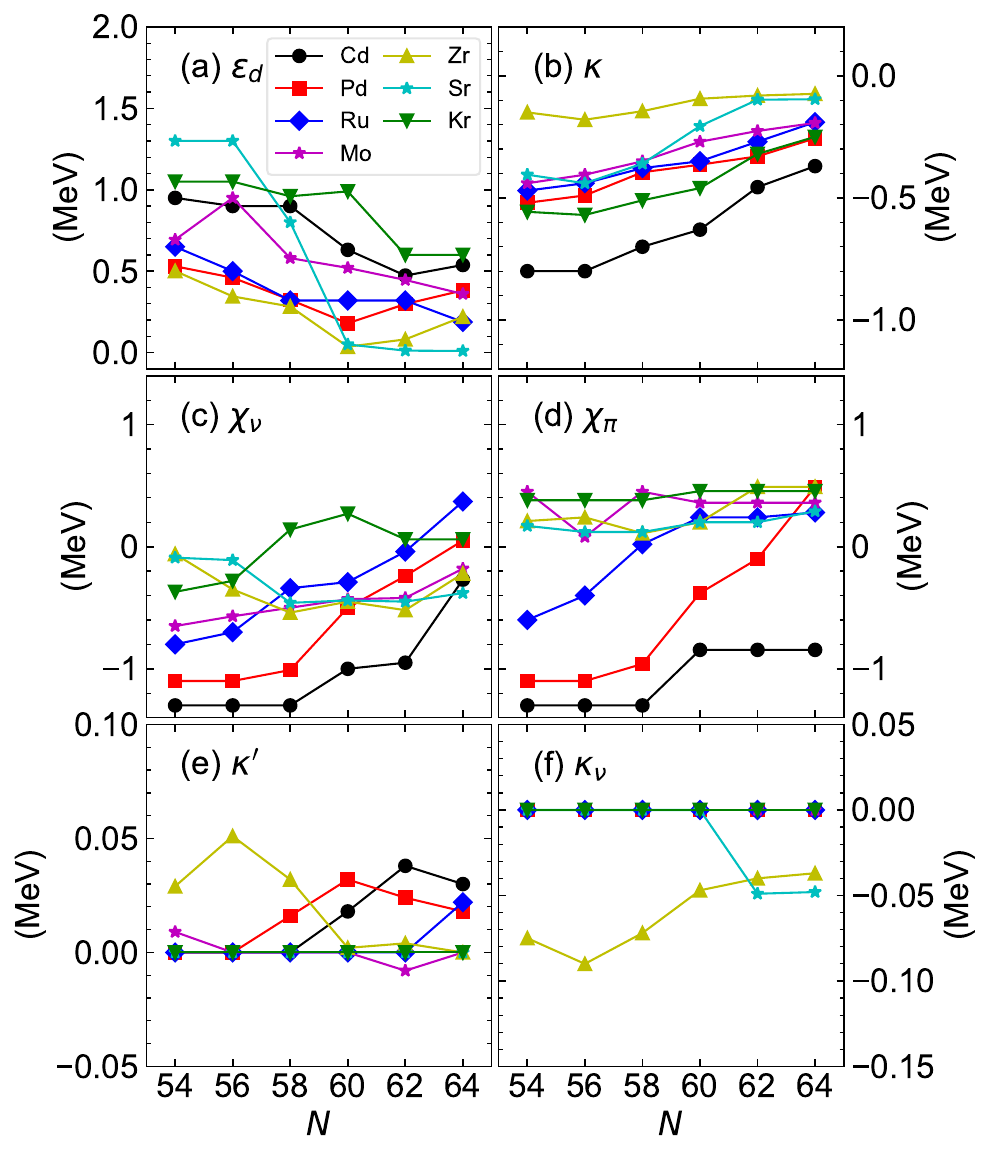}
\caption{Adopted strength parameters of the IBM-2 
Hamiltonian for the even-even nuclei studied in 
the present work.}
\label{fig:para-even}
\end{center}
\end{figure}

%
\begin{figure}
\begin{center}
\includegraphics[width=\linewidth]{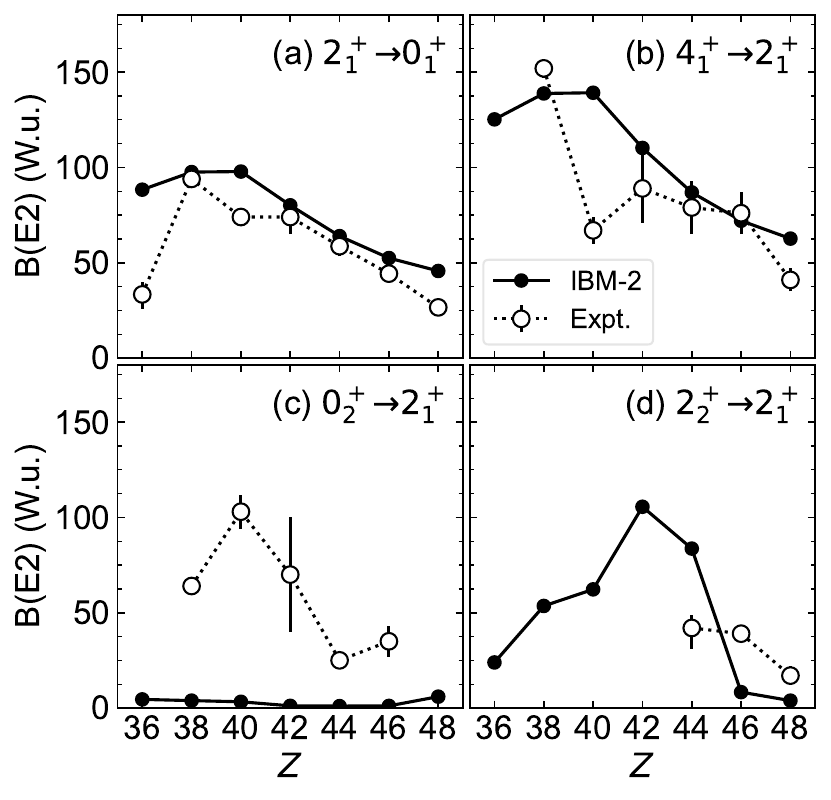}
\caption{Calculated and experimental \cite{data} 
(a) $B(E2;2^+_1 \to 0^+_1)$, (b) $B(E2;4^+_1 \to 2^+_1)$, 
(c) $B(E2;0^+_2 \to 2^+_1)$, and (d) $B(E2;2^+_2 \to 2^+_1)$ 
transition strengths for the even-even $N=60$ isotones 
as functions of the proton number $Z$.}
\label{fig:ee-em-n60}
\end{center}
\end{figure}

\subsection{IBM-2 results for the even-even nuclei}

Figure~\ref{fig:ee-level} depicts the excitation 
energies of the $2^+_1$, $4^+_1$, $0^+_2$, and $2^+_2$ 
states of the even-even Kr, Sr, Zr, Mo, Ru, Pd, 
and Cd nuclei with $54 \leqslant N \leqslant 64$ 
calculated with the mapped IBM-2. 
Note that the results for the Zr isotopes 
have already been presented in Ref.~\cite{nomura2020zr}, 
but they are included in the plot for 
the sake of completeness. 
One could observe in Fig.~\ref{fig:ee-level} 
that the mapped IBM-2 gives 
a reasonable description of the observed 
$2^+_1$ and $4^+_1$ excitation energies 
for all the isotopic chains, 
except for the Zr one. 
In many cases, the $0^+_2$ and $2^+_2$ energy levels are 
overestimated for nearly spherical nuclei 
that are close to the $N=50$ magic number, 
for which the IBM description 
in general becomes less reliable. 
For the Sr and Zr isotopes, the calculated 
low-lying levels exhibit a 
rapid decrease in energy starting from 
$N=56$ to 60. This behavior of the levels 
can be interpreted as a signature of the 
shape phase transition 
from the nearly spherical to deformed 
configurations. 
The high-lying $2^+_1$ energy level 
observed for $^{96}$Zr is due to 
the filling of the neutron $d_{5/2}$ subshell closure. 
As addressed in Ref.~\cite{nomura2020zr}, 
the major reason why the mapped IBM-2 is not able to 
reproduce the level structure of the Zr isotopes 
in the transitional regions, i.e., $N=56$ and 58, 
is that the RHB PESs for these nuclei suggest 
strong deformation and the resultant IBM-2 energy 
levels are rather compressed. 
In many of the previous mapped IBM-2 calculations, 
the $0^+_2$ excitation energies have been quite often 
overestimated. 
Such a problem could be explained in part by 
the fact that the underlying EDF calculation 
generally suggests a too large deformation 
and one has to choose the quadrupole-quadrupole 
interaction strength $\kappa$ that is 
unexpectedly larger in magnitude 
than those which have been 
often used in phenomenological 
IBM-2 fitting calculations.

Figure~\ref{fig:para-even} shows 
evolution of the derived IBM-2 parameters 
as functions of $N$ 
employed in the present calculation. 
One can find some correlations between 
the behaviors of the low-lying levels 
and those of the IBM-2 parameters. 
In Fig.~\ref{fig:para-even}(a), 
for instance, 
the decrease with $N$ 
of the single $d$ boson energy, $\epsilon_d$, 
indicates development of 
quadrupole deformation. 
The decrease in magnitude of the 
parameter $\kappa$ with $N$ is also a signature 
of increasing quadrupole collectivity 
[Fig.~\ref{fig:para-even}(b)]. 
The average of the parameters, 
$\chi\equiv (\chi_\nu+\chi_\pi)/2$, and 
its sign reflect whether 
the nucleus is prolate ($\chi<0$) 
or oblate ($\chi>0$) deformed in the 
SCMF calculations. 
For many of the nuclei, 
particularly in the Kr, Sr, Zr, and Mo isotopes, 
the average $\chi$ is close to zero, 
indicating the $\gamma$ softness that 
is found in the corresponding PESs (see, Fig.~\ref{fig:pes}). 
The $\hat L \cdot \hat L$ term is 
considered only for those nuclei for which 
the IB moment of inertia is calculated 
to be appreciable, i.e., 
approximately larger than 10. 
Otherwise, the strength parameter of this term 
is simply set as $\kappa'=0$.

Figure~\ref{fig:ee-em-n60} shows the 
calculated $B(E2)$ values in Weisskopf units (W.u.) 
for the even-even $N=60$ isotones. 
The mapped IBM-2 provides overall a 
reasonable quantitative and qualitative description 
of the $B(E2; 2^+_1 \to 0^+_1)$ [Fig.~\ref{fig:ee-em-n60}(a)]
and $B(E2; 4^+_1 \to 2^+_1)$ [Fig.~\ref{fig:ee-em-n60}(b)] transition 
strengths, even though the common 
boson effective charges are used for all 
the nuclei considered. 
An exception is perhaps the 
$B(E2; 4^+_1 \to 2^+_1)$ value for $^{100}$Zr, 
for which the IBM-2 gives twice as large as 
that of the experimental data. 
However, the observed $B(E2; 4^+_1 \to 2^+_1)$ 
value is 67$\pm 7$ W.u., which is  
smaller than, or in the same 
order of magnitude as, 
the $B(E2; 2^+_1 \to 0^+_1)$ value 
of 74$\pm 4$ W.u. 
This systematic would not be reproduced 
within the present IBM-2 model space 
consisting of a single configuration, 
giving rise to only the 
$B(E2; 4^+_1 \to 2^+_1)$ rate that is 
greater than the $B(E2; 2^+_1 \to 0^+_1)$ one.

As seen from Fig.~\ref{fig:ee-em-n60}(c) 
the observed $B(E2; 0^+_2 \to 2^+_1)$ rates 
of the $N=60$ isotones are generally 
large, that is, of the orders of $>20$ W.u. 
The large $B(E2; 0^+_2 \to 2^+_1)$ values 
are often considered a signature of 
strong shape mixing or shape coexistence. 
Particularly large $B(E2; 0^+_2 \to 2^+_1)$ values 
of $> 60$ W.u. that are found experimentally 
for the $^{98}$Sr, $^{100}$Zr, and $^{102}$Mo 
are considered to be a consequence 
of strong shape mixing in these nuclei. 
On the contrary, the mapped IBM-2 
suggests too small $B(E2; 0^+_2 \to 2^+_1)$ values 
for all the $N=60$ isotones. 
The vanishing $B(E2; 0^+_2 \to 2^+_1)$ 
rate implies that the mixing between the ground state 
and the $0^+_2$ states is not properly 
accounted for in the present calculation. 
A possible remedy for this inconsistency 
would be to include in the IBM-2 
the configuration mixing, since this extension 
often works in reproducing strong 
$B(E2; 0^+_2 \to 2^+_1)$ rates for those nuclei 
where shape coexistence is suggested to occur. 
Another major cause of the inconsistency could be 
attributed to the SCMF PESs calculated for 
these $N=60$ isotones, which generally suggest 
a too strong deformation. 
As a consequence, the mapped IBM-2 ends up  
giving a more rotational structure than 
expected, exhibiting a weak $0^+_2 \to 2^+_1$ transition. 

As one sees in Fig.~\ref{fig:ee-em-n60}(d), 
enhanced $B(E2; 2^+_2 \to 2^+_1)$ transition rates 
are predicted in the mapped IBM-2 calculation. 
This transition is often considered 
an indicator of the $\gamma$ softness, which 
is indeed shown to be most significant 
around $^{102}$Mo and $^{104}$Ru in the 
corresponding SCMF PESs (see Fig.~\ref{fig:pes}).

%
\begin{figure*}
\begin{center}
\includegraphics[width=0.49\linewidth]{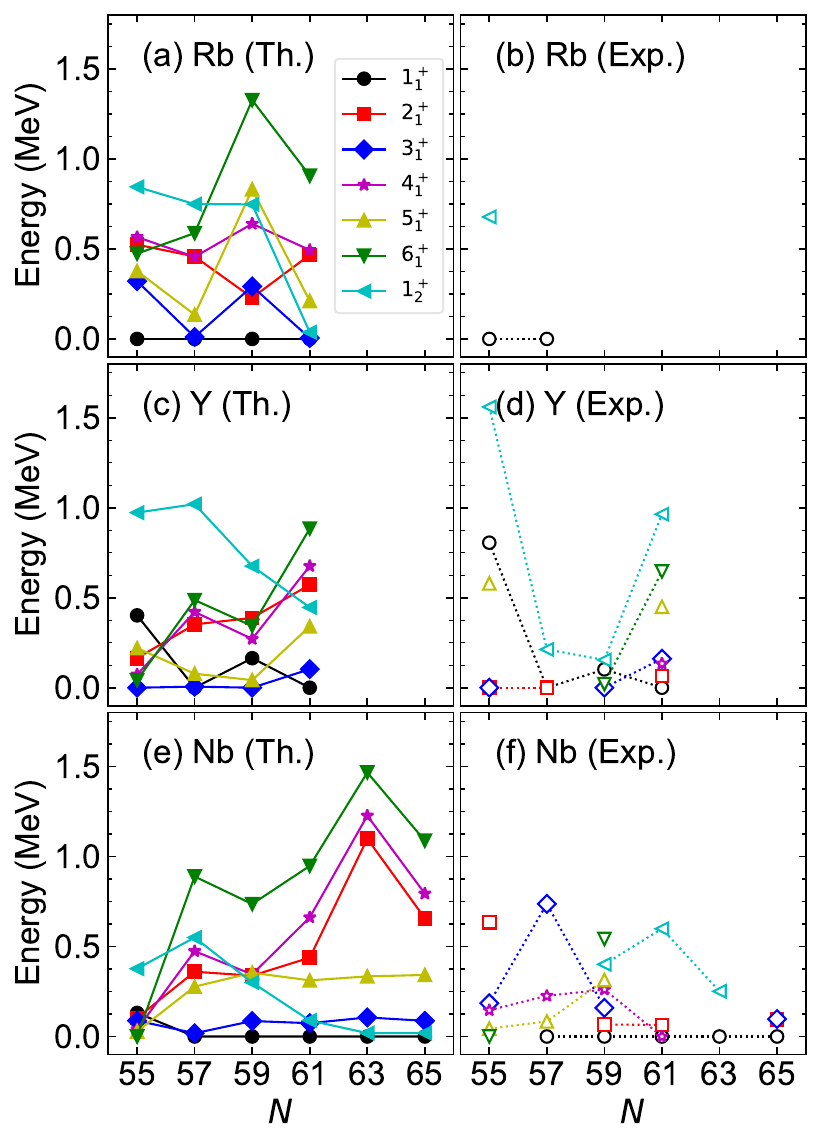}
\includegraphics[width=0.49\linewidth]{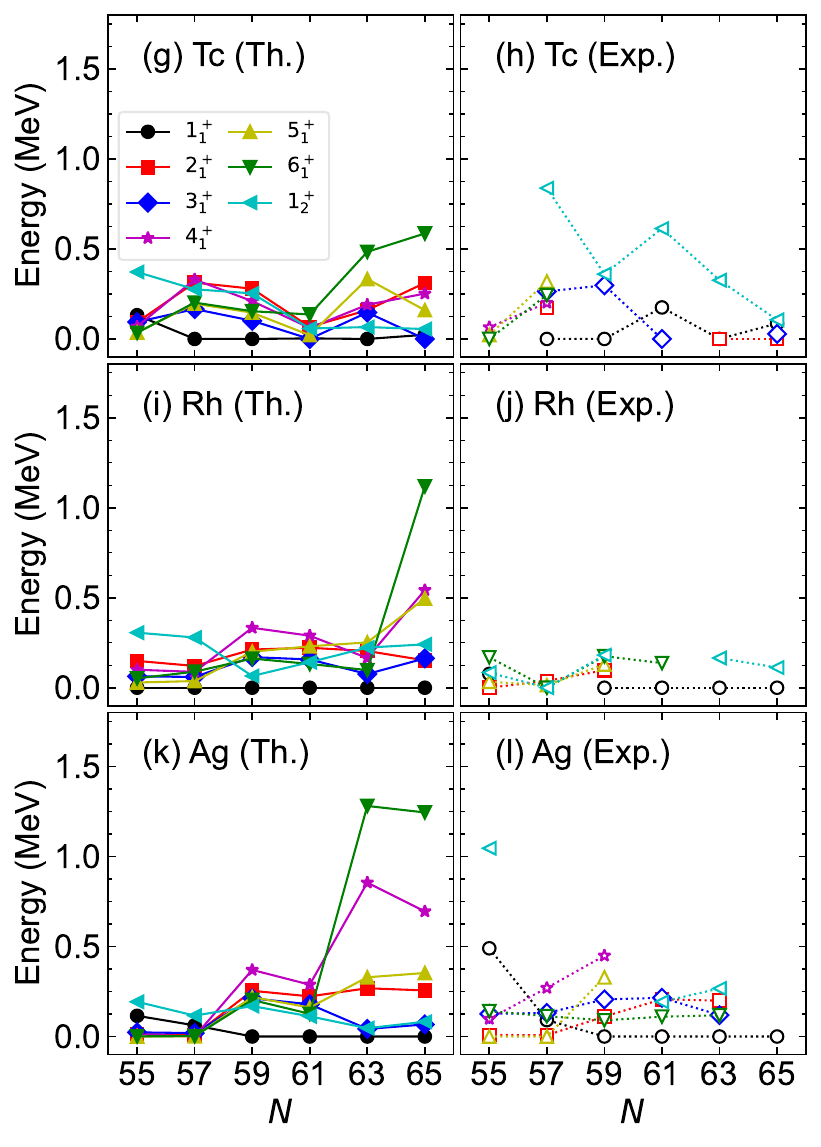}
\caption{Calculated and experimental 
excitation energies of the low-spin 
positive-parity states of the odd-odd nuclei.}
\label{fig:level-doo}
\end{center}
\end{figure*}

\begin{figure}
\begin{center}
\includegraphics[width=\linewidth]{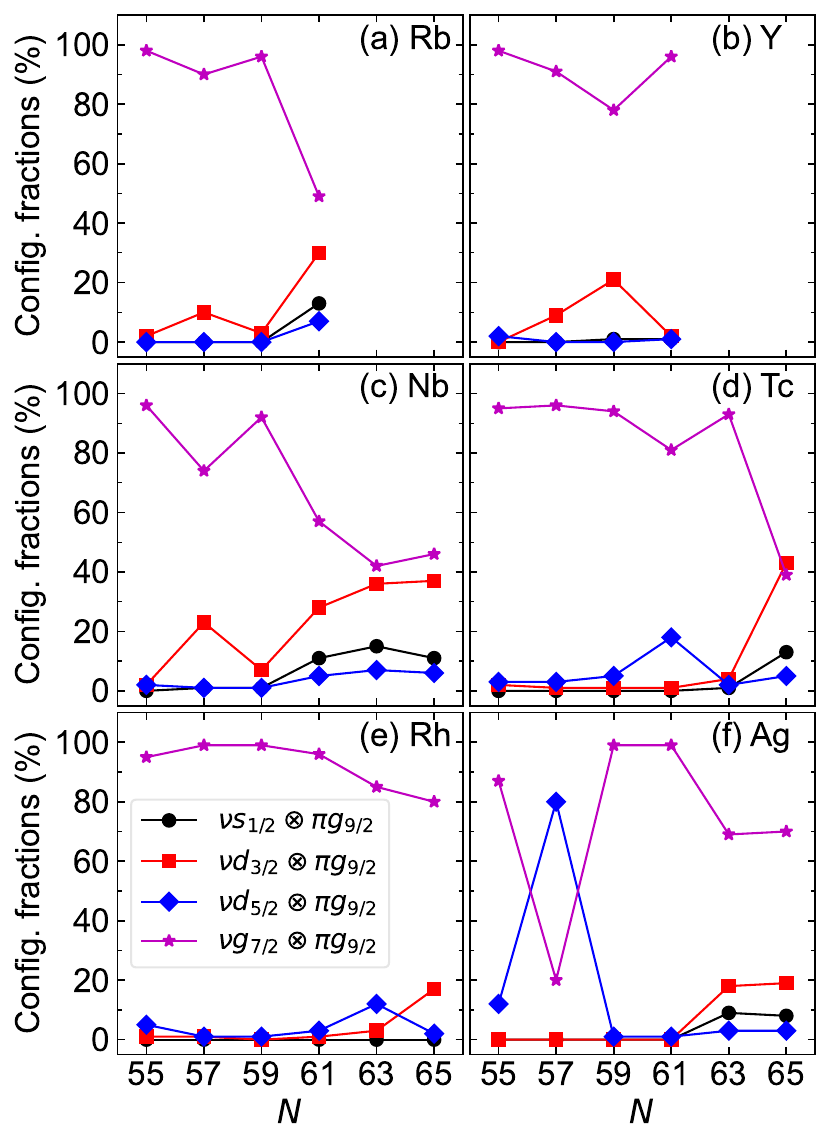}
\caption{Fractions in percent of the configurations 
in terms of the pair components 
$[\nu s_{1/2}\otimes \pi g_{9/2}]^{(J)}$, 
$[\nu d_{3/2}\otimes \pi g_{9/2}]^{(J)}$, 
$[\nu d_{5/2}\otimes \pi g_{9/2}]^{(J)}$, 
and 
$[\nu g_{7/2}\otimes \pi g_{9/2}]^{(J)}$ 
that are coupled to even-even boson 
cores in the IBFFM-2 wave function of the 
$1^+_1$ state for the studied odd-odd nuclei.}
\label{fig:config-doo_1+}
\end{center}
\end{figure}

\subsection{IBFFM-2 results for the odd-odd nuclei}

The calculated excitation spectra of the 
low-lying positive-parity states of 
the odd-odd Rb, Y, Nb, Tc, Rh, and Ag nuclei are 
presented in Fig.~\ref{fig:level-doo}, 
in comparison 
with the available experimental data \cite{data}. 
For Rb and Y isotopes calculations 
are made only for those nuclei for which 
experimental information is available or 
the same strength parameters of the 
IBFFM-2 Hamiltonian as those fitted to 
the available data on neighboring nuclei 
are used (for Rb). 
Theoretical excitation energies 
for $^{100}$Rb and $^{102}$Rb 
are therefore not shown in the figure. 
Calculated energies are also not 
shown for $^{102}$Y and $^{104}$Y, 
which is because the present version 
of the IBFFM-2 code is unable 
to handle the dimension of the 
IBFFM-2 Hamiltonian matrices for 
these nuclei.

One observes, in Fig.~\ref{fig:level-doo}, 
rapid evolution of energy levels 
at particular isotopes within the range $N=57-61$, 
which is in many cases accompanied by the change 
of the ground-state spin. 
The structural evolution is 
most clearly seen, both theoretically and 
experimentally, in the Y isotopic 
chain, and reflects the shape phase transition 
at $N \approx 60$ in the even-even boson Sr core 
[see Fig.~\ref{fig:ee-level}(b)]. 
For the Nb, Tc, Rh, and Ag 
isotopic chains, the majority of the isotopes 
with $N\geqslant 59$ have the $1^+$ state 
as the lowest-energy positive-parity state, 
while near the neutron shell closure 
$N=50$ those states with spin higher than $1^+$ 
become the ground state.

Table~\ref{tab:para-doo} lists the adopted 
strength parameters for the IBFFM-2 
Hamiltonian, i.e., those for the boson-fermion 
interactions ($\Gamma_\nu$, $\Gamma_\pi$, 
$\Lambda_\nu$, $\Lambda_\pi$, $A_\nu$, and $A_\pi$) 
and for the residual fermion-fermion 
interactions.
In the present calculations, the $\delta$-type 
and tensor interactions turn out to be 
most important to reproduce the 
low-energy spectra of odd-odd nuclei. 
Fixed values are used for the strength 
parameter for the $\delta$ term, $\vd=-0.08$ MeV, 
and that for the spin-spin-$\delta$ term 
is set to zero, $\vssd=0$ MeV. 
The spin-spin  
term is also assumed to be zero, 
but is introduced specifically for the 
$^{98}$Tc and $^{96}$Nb nuclei, with 
the corresponding strengths being 
$\vsss=0.1$ MeV for both nuclei, 
in order to reproduce the ground-state 
spin of $6^+$. 
The energy of the $1^+$ states 
turns out to be quite 
sensitive to the tensor interaction 
strength, $\vt$. 
As seen in Table~\ref{tab:para-doo}, 
the adopted $\vt$ strength indeed varies from 
one nucleus to another, so 
that the $1^+$ state should be the ground state 
in many of the nuclei.

%
\begin{table}
\caption{\label{tab:para-doo}
Adopted strength parameters (in MeV units) 
for the boson-fermion 
interactions, and fermion-fermion interactions 
in the IBFFM-2 Hamiltonian describing 
the considered odd-odd nuclei. 
The fixed values, $\vd=-0.08$ MeV, 
$\vssd=0.0$ MeV, are employed for the 
$\delta$ and spin-spin-$\delta$ terms, 
respectively. The spin-spin interaction 
strength, $\vsss$, is chosen to be zero 
for all nuclei, but for $\vsss=0.1$ MeV 
for $^{98}$Tc and $^{96}$Nb. 
}
 \begin{center}
 \begin{ruledtabular}
  \begin{tabular}{lccccccc}
Nucleus & $\Gamma_{\nu}$ & $\Gamma_{\pi}$ & $\Lambda_\nu$ & $\Lambda_\pi$ & $A_\nu$ & $A_\pi$ & $\vt$ \\
\hline
$^{102}$Ag & $0.30$ & $0.30$ & $0.70$ & $1.70$& $-0.50$ & $-2.00$ & $0.000$ \\
$^{104}$Ag & $0.30$ & $0.30$ & $0.70$ & $1.70$& $-0.20$ & $-0.70$ & $0.000$ \\
$^{106}$Ag & $0.30$ & $0.30$ & $0.70$ & $1.40$& $-0.00$ & $-0.70$ & $0.060$ \\
$^{108}$Ag & $0.30$ & $0.30$ & $0.70$ & $1.30$& $-0.00$ & $-0.00$ & $0.055$ \\
$^{110}$Ag & $0.10$ & $0.30$ & $3.00$ & $1.30$& $-0.20$ & $-0.80$ & $0.500$ \\
$^{112}$Ag & $0.10$ & $0.30$ & $2.60$ & $1.00$& $-0.30$ & $-0.00$ & $0.700$ \\
$^{100}$Rh & $0.30$ & $0.30$ & $0.60$ & $1.00$& $-0.00$ & $-0.80$ & $0.030$ \\
$^{102}$Rh & $0.30$ & $0.30$ & $0.20$ & $0.35$& $-0.00$ & $-0.00$ & $0.020$ \\
$^{104}$Rh & $0.30$ & $0.30$ & $0.20$ & $1.00$& $-0.00$ & $-0.80$ & $0.055$ \\
$^{106}$Rh & $0.30$ & $0.30$ & $0.00$ & $1.20$& $-0.00$ & $-0.00$ & $0.050$ \\
$^{108}$Rh & $1.50$ & $0.30$ & $0.70$ & $1.00$& $-0.20$ & $-0.40$ & $0.050$ \\
$^{110}$Rh & $1.50$ & $0.30$ & $0.10$ & $1.00$& $-0.50$ & $-0.50$ & $0.400$ \\
$^{98}$Tc & $0.30$ & $0.30$ & $0.40$ & $0.70$& $-0.00$ & $-0.00$ & $0.000$ \\
$^{100}$Tc & $0.30$ & $0.30$ & $0.35$ & $0.90$& $-0.00$ & $-0.00$ & $0.050$ \\
$^{102}$Tc & $0.30$ & $0.30$ & $0.44$ & $0.70$& $-0.00$ & $-0.00$ & $0.040$ \\
$^{104}$Tc & $0.30$ & $0.30$ & $0.60$ & $2.60$& $-0.40$ & $-3.00$ & $0.030$ \\
$^{106}$Tc & $0.30$ & $0.30$ & $0.30$ & $2.10$& $-0.00$ & $-3.00$ & $0.150$ \\
$^{108}$Tc & $0.30$ & $0.10$ & $0.90$ & $2.20$& $-1.20$ & $-0.00$ & $0.500$ \\
$^{96}$Nb & $0.30$ & $0.30$ & $0.40$ & $0.90$& $-0.00$ & $-0.50$ & $-0.000$ \\
$^{98}$Nb & $1.50$ & $0.10$ & $0.80$ & $0.00$& $-1.20$ & $-0.00$ & $0.280$ \\
$^{100}$Nb & $1.50$ & $0.30$ & $0.50$ & $0.20$& $-1.40$ & $-0.80$ & $0.500$ \\
$^{102}$Nb & $1.50$ & $0.10$ & $0.90$ & $3.80$& $-0.90$ & $-2.00$ & $0.500$ \\
$^{104}$Nb & $1.50$ & $0.10$ & $1.55$ & $3.80$& $-0.90$ & $-2.00$ & $0.800$ \\
$^{106}$Nb & $1.50$ & $0.10$ & $1.00$ & $3.80$& $-0.30$ & $-2.00$ & $0.800$ \\
$^{94}$Y & $0.10$ & $1.00$ & $0.40$ & $0.00$& $-1.00$ & $-0.40$ & $0.000$ \\
$^{96}$Y & $0.50$ & $1.00$ & $0.10$ & $0.00$& $-0.00$ & $-0.50$ & $0.060$ \\
$^{98}$Y & $0.50$ & $1.00$ & $0.10$ & $0.00$& $-0.00$ & $-0.00$ & $0.040$ \\
$^{100}$Y & $0.50$ & $0.50$ & $0.80$ & $12.00$& $-0.80$ & $-0.00$ & $2.000$ \\
$^{92}$Rb & $0.30$ & $0.50$ & $0.40$ & $0.00$& $-1.00$ & $-0.30$ & $0.080$ \\
$^{94}$Rb & $0.70$ & $0.50$ & $0.20$ & $0.00$& $-2.00$ & $-1.00$ & $0.100$ \\
$^{96}$Rb & $0.70$ & $0.50$ & $0.00$ & $0.00$& $-0.00$ & $-2.00$ & $0.500$ \\
$^{98}$Rb & $0.30$ & $0.10$ & $2.20$ & $34.00$& $-5.00$ & $-5.00$ & $0.500$ \\
 \end{tabular}
 \end{ruledtabular}
 \end{center}
\end{table}

\begin{table}
\caption{\label{tab:doo-em}
Comparison of the calculated 
and experimental \cite{data,stone2005} 
$B(E2)$ (in W.u.) and $B(M1)$ (in $\mu^2_N$) 
transitions, $Q(I)$ (in $e$b), and 
$\mu(I)$ (in $\mu_N$) moments for 
the odd-odd nuclei considered in this study. 
}
 \begin{center}
 \begin{ruledtabular}
  \begin{tabular}{lccc}
Nucleus 
& Property & IBFFM-2 & Experiment \\
\hline
$^{96}$Rb
& $\mu(2^+_{1})$ & 2.42 & $+1.466\pm0.002$ \\
& $Q(2^+_{1})$ & $0.48$ & $+0.25\pm0.06$ \\
$^{96}$Nb
& $\mu(6^+_{1})$ & 4.71 & $4.976\pm0.004$ \\
$^{100}$Nb
& $B(M1;1^+_{2}\to1^+_{1})$ & $0.25$ & $0.0038^{+0.0019}_{-0.0010}$ \\
$^{102}$Ag
& $B(E2;6^+_{1}\to5^+_{1})$ & $0.64$ & $32\pm16$ \\
& $B(M1;6^+_{1}\to5^+_{1})$ & $1.18$ & $0.0011\pm0.0004$ \\
& $\mu(5^+_{1})$ & 4.38 & $+4.6\pm0.7$ \\
& $\mu(2^+_{1})$ & 5.43 & $+4.1\pm0.3$ \\
& $\mu(7^+_{1})$ & 4.52 & $4.6\pm0.3$ \\
$^{104}$Ag
& $B(M1;6^+_{1}\to5^+_{1})$ & $0.12$ & $>0.0089$ \\
& $B(M1;7^+_{1}\to6^+_{1})$ & $0.03$ & $0.0085\pm0.0020$ \\
& $B(E2;7^+_{1}\to6^+_{1})$ & $28.96$ & $(1.2\pm1.1)\times10^2$ \\
& $B(E2;7^+_{1}\to5^+_{1})$ & $4.37$ & $1.4\pm0.3$ \\
& $\mu(5^+_{1})$ & 3.93 & $3.917\pm0.008$ \\
& $\mu(2^+_{1})$ & 4.88 & $+3.7\pm0.2$ \\
$^{106}$Ag
& $\mu(1^+_{1})$ & 3.03 & $+2.9\pm0.2$ \\
& $Q(6^+_{1})$ & $0.73$ & $+1.08\pm0.06$ \\
& $\mu(6^+_{1})$ & 4.08 & $(+)3.705\pm0.004$ \\
$^{108}$Ag
& $B(E2;1^+_{2}\to1^+_{1})$ & $30.61$ & $>3.1$ \\
& $B(E2;3^+_{1}\to1^+_{1})$ & $7.33$ & $0.85\pm0.08$ \\
& $B(M1;2^+_{1}\to1^+_{1})$ & $0.02$ & $>0.012$ \\
& $B(M1;1^+_{2}\to1^+_{1})$ & $0.00$ & $>0.0056$ \\
& $\mu(1^+_{1})$ & 3.00 & $2.6884\pm0.0007$ \\
& $Q(6^+_{1})$ & $0.89$ & $+1.32\pm0.07$ \\
& $\mu(3^+_{1})$ & 3.33 & $3.888\pm0.015$ \\
& $\mu(6^+_{1})$ & 3.93 & $3.58\pm0.02$ \\
$^{110}$Ag
& $Q(1^+_{1})$ & $0.06$ & $0.24\pm0.12$ \\
& $\mu(1^+_{1})$ & 3.05 & $2.7271\pm0.0008$ \\
& $Q(6^+_{1})$ & $-0.03$ & $+1.41\pm0.10$ \\
& $\mu(6^+_{1})$ & 4.19 & $+3.607\pm0.004$ \\
& $\mu(3^+_{1})$ & 3.53 & $+3.77\pm0.03$ \\
  \end{tabular}
 \end{ruledtabular}
 \end{center}
\end{table}

Figure~\ref{fig:config-doo_1+} exhibits 
fractions of the pair components denoted 
$[\nu s_{1/2}\otimes \pi g_{9/2}]^{(J)}$, 
$[\nu d_{3/2}\otimes \pi g_{9/2}]^{(J)}$, 
$[\nu d_{5/2}\otimes \pi g_{9/2}]^{(J)}$,
and 
$[\nu g_{7/2}\otimes \pi g_{9/2}]^{(J)}$
in the wave functions of the $1^+_1$ state 
of the odd-odd nuclei. In most of the cases 
shown in the figure, the configuration 
of the $[\nu g_{7/2}\otimes \pi g_{9/2}]^{(J)}$ 
pairs coupled to the even-even boson 
cores predominates the $1^+_1$ wave functions 
typically for those nuclei with 
the neutron numbers $N\leqslant 59$. 
For heavier isotopes with $N$ larger than 59, 
other pair components start to play a role, 
especially 
the $[\nu d_{3/2}\otimes \pi g_{9/2}]^{(J)}$ 
ones. 
The $[\nu s_{1/2}\otimes \pi g_{9/2}]^{(J)}$ 
pairs do not appear to play an important 
role in the $1^+_1$ state of all the nuclei 
considered. 
The change in the composition of the 
wave function for the odd-odd nuclei 
reflects the shape phase transitions 
in the even-even core nuclei and, 
as shown later, influences the predicted 
systematic of $\beta$-decay properties.

Table~\ref{tab:doo-em} compares the calculated 
and experimental $B(E2)$ and $B(M1)$ transition 
rates, and electric quadrupole $Q(I)$ 
and magnetic dipole $\mu(I)$ moments. 
One notices that the present IBFFM-2 generally gives 
a reasonable description of the $Q(I)$ and 
$\mu(I)$ moments including sign. 
Limited experimental data are available to 
compare for the $B(E2)$ and $B(M1)$ transition 
rates. 
Some deviations are present between the calculated and 
experimental data for the transition rates, 
which could be understood in terms of 
the different nature of the IBFFM-2 wave functions 
for the initial and final states. 
In $^{102}$Ag, for example, 
the present IBFFM-2 calculation underestimates the 
observed $B(E2)$ rates, but overestimates 
the $B(M1)$ ones. 
The dominant pair component 
in the $5^+_1$ ground state for $^{102}$Ag 
is here suggested to be of the type 
$[\nu d_{5/2}\otimes \pi g_{9/2}]^{(J=5)}$, 
which accounts for 95 \% of the corresponding 
wave function. 
On the other hand, the $6^+_1$ state 
for the same nucleus is suggested to be dominated 
by the $[\nu d_{5/2}\otimes \pi g_{9/2}]^{(J=6)}$ 
(87 \%) and 
$[\nu d_{5/2}\otimes \pi g_{9/2}]^{(J=7)}$
(10 \%) pair components. 
For both the $E2$ and $M1$ 
$6^+_1 \to 5^+_1$ transitions, 
the fermionic part of the transition operators, 
$\hat T^{(E2/M1)}_{\rm F}$ 
[Eqs.~(\ref{eq:e2}) and (\ref{eq:m1})], 
are found to make major contributions 
to the transition matrix elements. 
The fermion contribution seems to be less significant 
in the $E2$ matrix element, but makes 
a substantial contribution to the $M1$ matrix element. 
These observations may help in understanding 
the deviations in the transition properties 
shown in Table~\ref{tab:doo-em}. 

In general, however, the experimental information 
about the electromagnetic transitions is 
still so scarce in the odd-odd nuclei in 
the considered mass region that it is rather 
hard to draw a concrete conclusion 
about the performance of the IBFFM-2 
in computing these observables.

\section{$\beta$ decay properties\label{sec:beta}}

%
\begin{figure*}
\begin{center}
\includegraphics[width=.49\linewidth]{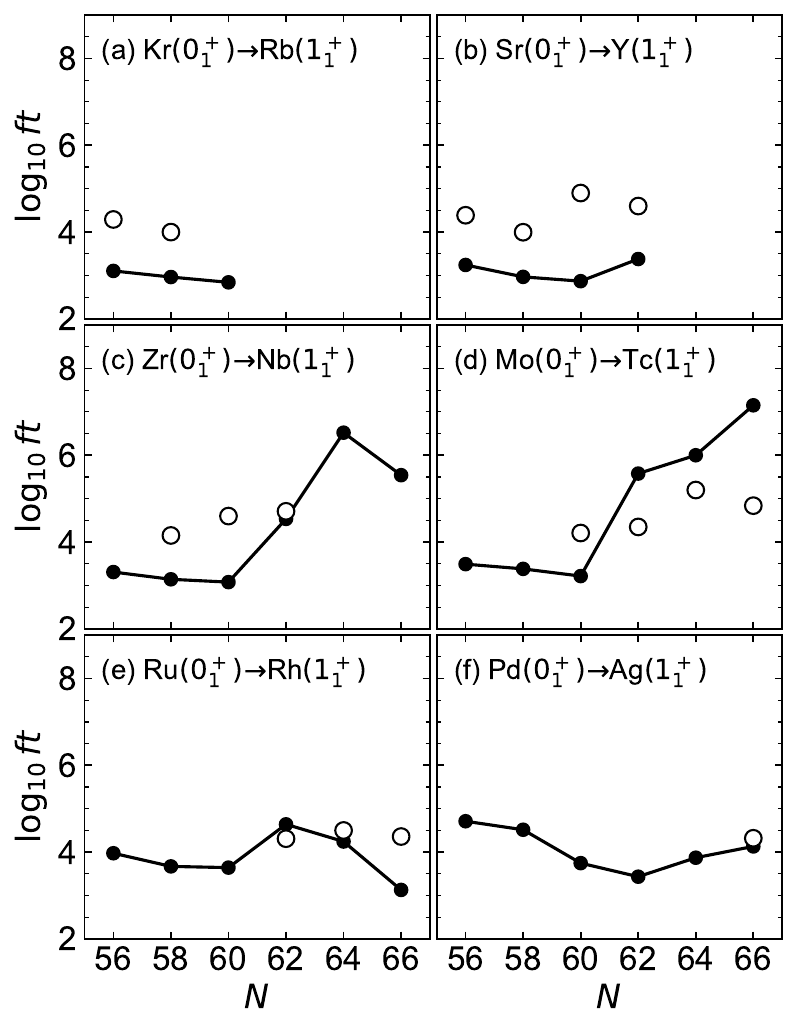}
\includegraphics[width=.49\linewidth]{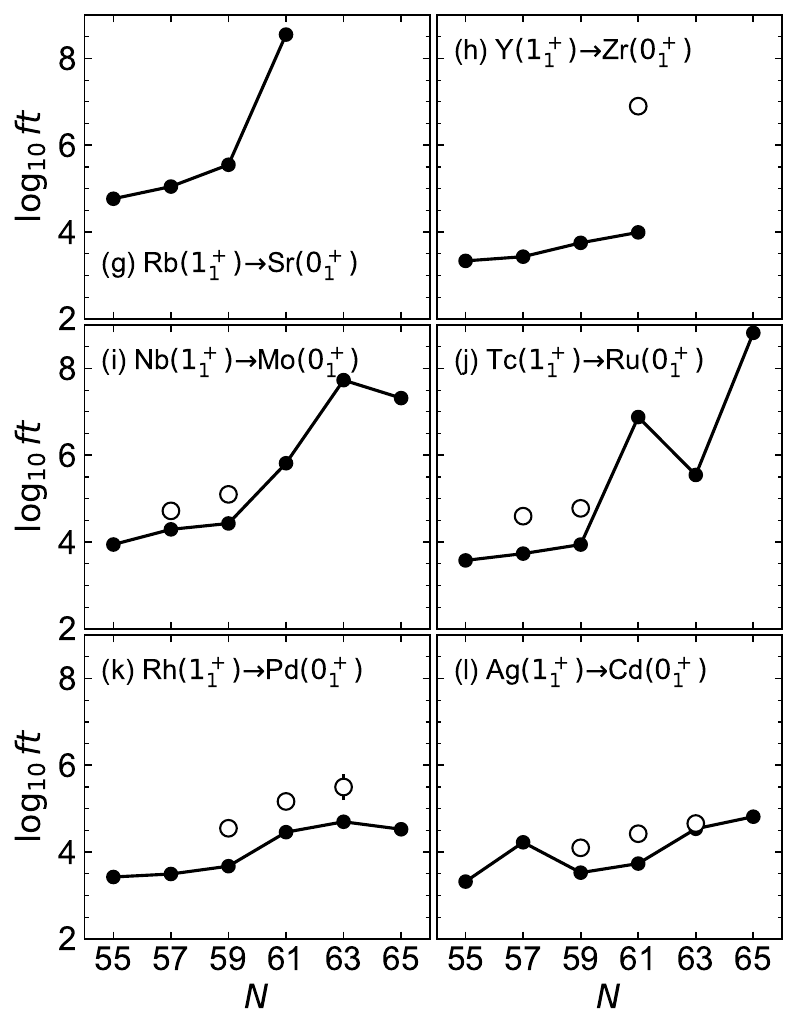}
\caption{The $\ft$ values for [(a)-(f)] the $\btm$ decay 
of the $0^+_1$ ground state of even-even nuclei 
into the $1^+_1$ state of odd-odd nuclei, 
and [(g)-(l)] the $\btm$ 
decay of the $1^+_1$ state 
of odd-odd nuclei into the $0^+_1$ state of 
even-even nuclei. 
Calculated $\ft$ values are represented by 
solid circles connected by lines, and 
open circles denote the experimental values \cite{data}.}
\label{fig:ft-even}
\end{center}
\end{figure*}

\subsection{$\ft$ values\label{sec:ft}}

The wave functions of the initial and 
final states resulting from the IBM-2 
and IBFFM-2 Hamiltonians are used to calculate 
the matrix elements of the GT operator (\ref{eq:ogt}), 
denoted by $\mgt$, which are necessary  
to obtain $ft$ values in seconds: 
\begin{eqnarray}
 \label{eq:ft}
ft=\frac{6163}{\left(\frac{\ga}{\gv}\right)^2 |\mgt|^2} \; .
\end{eqnarray}
$\ga=1.27$ and $\gv=1$ are the free values of 
the axial vector and vector coupling constants, 
respectively. 
In order to better describe the $\beta$ decay rate, 
effective $\ga$ factors are often considered 
by quenching the free value. 
As shown in the following, the effective $\ga$ 
values that could be estimated from the observed 
$\ft$ values are quite at variance with 
different nuclei; that is, in some cases 
the $\ga$ factor would need to be reduced, while 
in other cases it should be increased. 
For the sake of simplicity, the free value, 
$\ga=1.27$, is used throughout in the present study 
for calculating $ft$ values.

In Figs.~\ref{fig:ft-even}(a)--\ref{fig:ft-even}(f) 
the calculated $\ft$ values for the 
$\btm$ decay of the $0^+_1$ 
ground state of even-even nuclei into the $1^+_1$ 
state of the odd-odd nuclei are presented 
as functions of $N$ for each isotopic chain. 
One notices that the $0^+_1 \to 1^+_1$ 
$\btm$ decay $\ft$ values are systematically 
lower than the measured ones \cite{data}, 
mainly in those nuclei located before the 
shape phase transitions, i.e., 
$N\leqslant 60$, 
and with lower proton numbers, 
i.e., Kr, Sr, and Zr ones. 
This finding indicates 
that an enormous amount of quenching would 
need to be made of the GT transition matrix elements. 
In the case of the $^{98}$Sr$(0^+_1)$ 
$\to$ $^{98}$Y$(1^+_1)$ decay, for instance, 
an effective $\ga$ factor that is 
one order of magnitude smaller than 
the free value $\ga=1.27$ would be 
required to reproduce the experimental $\ft$ 
value of $4.9\pm 0.1$ \cite{data}.

As often suggested 
(see, e.g., Ref.~\cite{yoshida2013}),  
the quenching of the GT matrix elements 
implies certain deficiencies of 
the employed nuclear structure models 
regarding, e.g., various model assumptions, 
parameters, and restricted configuration spaces. 
In the present case, indeed, 
the assumption of taking 
only the $g_{9/2}$ orbital 
for the proton single-particle space 
seems reasonable for 
those isotopes with $Z$ being 
near the $Z=50$ major shell closure, 
for which the $g_{9/2}$ orbital 
plays an predominant role, whereas 
for the low-$Z$ nuclei, which are rather 
close to the $Z=28$ major shell closure, 
some other single-particle states, such 
as those coming from outside 
of the $Z=28-50$ major shell, may play 
a role.

The calculated GT matrix element 
can be analyzed by decomposing it into 
components that are 
associated with different neutron-proton 
pair configurations. 
For the $^{98}$Sr$(0^+_1)$ 
$\to$ $^{98}$Y$(1^+_1)$ decay, for example, 
the dominant contribution to the 
GT transition comes from the 
matrix elements of the terms in 
the GT operator that are of the forms like 
$[a^\+_{\nu g_{7/2}}\times a^\+_{\pi g_{9/2}}]^{(1)}$
and 
$[\tilde d_{\nu} \times [a^\+_{\nu g_{7/2}}\times a^\+_{\pi g_{9/2}}]^{(L)}]^{(1)}$, which are calculated to be too large in magnitude, 
thus leading to the small $\ft$ values.

One can also observe in 
Figs.~\ref{fig:ft-even}(a)--\ref{fig:ft-even}(f) 
that for the $\btm$ decays of the Zr and Mo 
isotopes 
the calculated $\ft$ values exhibit a drastic 
increase around $N=60$. 
In the Mo isotopic chain, the predicted $\ft$ 
values are even larger than the measured 
ones for the heavier isotopes with 
$N \geqslant 62$. 
Here, the rapid increase of the $\ft$ values 
at $N\approx 60$ could be also explained by 
the dominance of the 
neutron-proton pair configuration 
in the IBFFM-2 $1^+_1$ wave functions. 
As mentioned above, for those nuclei with 
$N \leqslant 60$, the matrix elements 
of the terms that involve the 
pair configurations of the type  
$[\nu g_{7/2}\otimes \pi g_{9/2}]^{(J)}$ 
are shown to be large.
However, this does not seem to be the case for 
the nuclei with $N>60$, where in the corresponding 
$1^+_1$ wave functions some other 
pair configurations start to 
make appreciable contributions 
(see, Fig.~\ref{fig:config-doo_1+}). 
The different pair components 
are more or less fragmented in 
the GT matrix elements for the $N>60$ nuclei, 
and cancel each other to give rise to 
the rather small $\mgt$, or large $\ft$ value.

Given that the $\ft$ values increase sharply 
for the $N>60$ even-even nuclei, and that 
in some cases [e.g., the Mo decays in Fig.~\ref{fig:ft-even}(d)] 
overestimate the data, 
then the $\ga$ factor would need to be 
rather increased for these neutron-rich isotopes, 
whereas a quenched $\ga$ factor would be required 
for the lighter isotopes with $N \leqslant 60$. 
As mentioned earlier, 
an immediate cause of this peculiarity 
is the change with $N$ 
in the nature of the wave functions for the 
odd-odd nuclei. 
The nuclear wave functions should 
be, however, sensitive to the choice of the 
single-particle spaces, Hamiltonian 
parameters, and single-particle properties 
($\tilde\epsilon_{\jr}$ and $v^2_{\jr}$), 
which are determined largely 
by the underlying EDF-SCMF calculation, 
and are at variance with different EDFs. 
This point is pursued further in Sec.~\ref{sec:edf}. 
The sensitivity of the mapped IBM-2 
prediction of the 
$\tnbb$-decay NMEs to the single-particle spaces 
and to the different EDFs has also been 
addressed in Ref.~\cite{nomura2022bb}.

As compared to the $0^+_1 \to 1^+_1$ $\btm$ decay 
of the lighter even-even nuclei with $N \leqslant 60$ 
and $Z<40$, variation of the $\ft$ 
values with $N$ occurs much more slowly 
for the Ru$(0^+_1)\to$Rh$(1^+_1)$ 
[Fig.~\ref{fig:ft-even}(e)] and 
Pd$(0^+_1)\to$Ag$(1^+_1)$ [Fig.~\ref{fig:ft-even}(f)] 
$\btm$ decays.

In a similar fashion, 
in Figs.~\ref{fig:ft-even}(g)--\ref{fig:ft-even}(l)
the predicted $\ft$ values for 
the $\btm$ decay of the odd-odd nuclei 
are compared with the experimental 
counterparts \cite{data}. 
Experimental data are not available for the 
Rb$(1^+_1)\to$Sr$(0^+_1)$ decay, and only 
the lower bound, $\ft>6.9$, of the 
unidentified $(1)^-$ state is known 
for the 
$^{100}$Y$(1^+_1)\to^{100}$Zr$(0^+_1)$ decay. 
A general remark is that in each isotopic 
chain the calculated 
$1^+_1 \to 0^+_1$ $\ft$ value increases 
as a function of $N$, consistently 
with the observed systematic, and 
further exhibits a rapid increase 
from $N=59$ to 61. 
The change is 
particularly significant for the 
Nb$(1^+_1)\to$Mo$(0^+_1)$ 
and Tc$(1^+_1)\to$Ru$(0^+_1)$ decays, 
while for the 
Rh$(1^+_1)\to$Pd$(0^+_1)$ and 
Ag$(1^+_1)\to$Cd$(0^+_1)$ ones 
the predicted $\ft$ values increase 
more slowly with $N$. 
In addition, the present calculation 
reproduces the measured $\ft$ values 
to a greater extent than in the case of the 
$\btm$ decay of the even-even nuclei.

%
\begin{table}
\caption{\label{tab:ft-btm}
Comparison of calculated and observed \cite{data} 
$\ft$ values for the $\btm$ decay of the 
even-mass nuclei including the $N=60$ 
isotones. 
}
 \begin{center}
 \begin{ruledtabular}
  \begin{tabular}{lccc}
& & \multicolumn{2}{c}{$\ft$} \\
\cline{3-4}
Decay & $I\to I'$ & Theo. & Exp. \\
\hline
$^{98}$Rb$\to^{98}$Sr
& $3^{+}_{1}\to2^{+}_{1}$ & 7.78 & 5.6 \\
& $3^{+}_{1}\to2^{+}_{2}$ & 9.95 & 6.2 \\
& $3^{+}_{1}\to2^{+}_{3}$ & 8.44 & 6.1 \\
& $3^{+}_{1}\to2^{+}_{4}$ & 8.05 & 5.5 \\
& $3^{+}_{1}\to4^{+}_{1}$ & 7.90 & 6.3 \\
$^{98}$Sr$\to^{98}$Y
& $0^{+}_{1}\to1^{+}_{1}$ & 2.87 & $4.9\pm0.1$ \\ 
& $0^{+}_{1}\to1^{+}_{2}$ & 4.38 & $4.4\pm0.1$ \\ 
& $0^{+}_{1}\to1^{+}_{3}$ & 4.82 & $5.5\pm0.1$ \\ 
& $0^{+}_{1}\to1^{+}_{4}$ & 4.20 & $5.6\pm0.1$ \\ 
$^{100}$Y$\to^{100}$Zr
& $4^{+}_{1}\to3^{+}_{1}$ & 4.22 & $>6.5$ \\
& $4^{+}_{1}\to4^{+}_{1}$ & 5.46 & $\approx6.0$ \\
& $4^{+}_{1}\to4^{+}_{2}$ & 4.54 & $>6.6$ \\
& $4^{+}_{1}\to4^{+}_{3}$ & 4.83 & $>6.9$ \\
$^{100}$Zr$\to^{100}$Nb
& $0^{+}_{1}\to1^{+}_{1}$ & 3.08 & $4.6\pm0.1$ \\ 
& $0^{+}_{1}\to1^{+}_{2}$ & 3.90 & $4.8\pm0.1$ \\ 
& $0^{+}_{1}\to1^{+}_{3}$ & 6.50 & $5.8$ \\ 
& $0^{+}_{1}\to1^{+}_{4}$ & 6.34 & $4.5\pm0.1$ \\ 
$^{102}$Nb$\to^{102}$Mo
& $1^{+}_{1}\to0^{+}_{1}$ & 5.82 & ${}$ \\
& $1^{+}_{1}\to0^{+}_{2}$ & 6.99 & ${}$ \\
& $1^{+}_{1}\to2^{+}_{1}$ & 6.72 & ${}$ \\
& $1^{+}_{1}\to2^{+}_{2}$ & 7.53 & ${}$ \\
& $4^{+}_{1}\to3^{+}_{1}$ & 6.45 & $6.21$ \\
& $4^{+}_{1}\to3^{+}_{2}$ & 6.20 & $4.86$ \\
& $4^{+}_{1}\to4^{+}_{1}$ & 7.51 & $6.44$ \\
& $4^{+}_{1}\to4^{+}_{2}$ & 7.45 & $6.31$ \\
$^{102}$Mo$\to^{102}$Tc
& $0^{+}_{1}\to1^{+}_{1}$ & 3.22 & $4.21\pm0.09$ \\ 
& $0^{+}_{1}\to1^{+}_{2}$ & 4.48 & $5.74\pm0.10$ \\ 
& $0^{+}_{1}\to1^{+}_{3}$ & 5.19 & $4.81\pm0.09$ \\ 
$^{104}$Tc$\to^{104}$Ru
& $3^{+}_{1}\to2^{+}_{1}$ & 6.22 & $8.10\pm0.07$ \\
& $3^{+}_{1}\to2^{+}_{2}$ & 5.63 & $8.14\pm0.09$ \\
& $3^{+}_{1}\to3^{+}_{1}$ & 7.83 & $8.18\pm0.10$ \\
& $3^{+}_{1}\to4^{+}_{1}$ & 6.91 & $8.34\pm0.09$ \\
$^{106}$Rh$\to^{106}$Pd
& $1^{+}_{1}\to0^{+}_{1}$ & 4.46 & $5.168\pm0.007$ \\
& $1^{+}_{1}\to0^{+}_{2}$ & 6.45 & $5.354\pm0.019$ \\
& $1^{+}_{1}\to0^{+}_{3}$ & 7.15 & $5.51\pm0.05$ \\
& $1^{+}_{1}\to0^{+}_{4}$ & 7.62 & $5.63\pm0.07$ \\
& $1^{+}_{1}\to2^{+}_{1}$ & 4.11 & $5.865\pm0.017$ \\
& $1^{+}_{1}\to2^{+}_{2}$ & 7.72 & $6.55\pm0.07$ \\
& $1^{+}_{1}\to2^{+}_{3}$ & 6.48 & $5.757\pm0.017$ \\
& $1^{+}_{1}\to2^{+}_{4}$ & 6.35 & $6.25\pm0.03$ \\
& $1^{+}_{1}\to2^{+}_{5}$ & 5.06 & $5.83\pm0.03$ \\
& $1^{+}_{1}\to2^{+}_{6}$ & 7.52 & $6.59\pm0.10$ \\
$^{108}$Ag$\to^{108}$Cd
& $1^{+}_{1}\to0^{+}_{1}$ & 3.74 & $4.425\pm0.009$ \\
& $1^{+}_{1}\to2^{+}_{1}$ & 6.06 & $5.35\pm0.03$ \\
 \end{tabular}
 \end{ruledtabular}
 \end{center}
\end{table}

%
\begin{table}
\caption{\label{tab:ft-ec}
Same as Table~\ref{tab:ft-btm}, but for 
the EC decay.
}
 \begin{center}
 \begin{ruledtabular}
  \begin{tabular}{lccc}
& & \multicolumn{2}{c}{$\ft$} \\
\cline{3-4}
Decay & $I\to I'$ & Theo. & Exp. \\
\hline
$^{104}$Rh$\to^{104}$Ru
& $1^{+}_{1}\to0^{+}_{1}$ & 4.12 & $4.32\pm0.11$ \\
& $1^{+}_{1}\to0^{+}_{2}$ & 4.93 & $5.15\pm0.18$ \\
& $1^{+}_{1}\to2^{+}_{1}$ & 6.26 & $5.42\pm0.06$ \\
$^{106}$Ag$\to^{106}$Pd
& $1^{+}_{1}\to0^{+}_{1}$ & 4.22 & $4.92$ \\
& $1^{+}_{1}\to0^{+}_{2}$ & 4.48 & $6.5$ \\
& $1^{+}_{1}\to0^{+}_{3}$ & 5.32 & $7.0$ \\
& $1^{+}_{1}\to0^{+}_{4}$ & 5.88 & $6.1$ \\
& $1^{+}_{1}\to0^{+}_{5}$ & 6.65 & $7.6$ \\
& $1^{+}_{1}\to0^{+}_{6}$ & 5.89 & $7.2$ \\
& $1^{+}_{1}\to0^{+}_{7}$ & 5.25 & $6.6$ \\
& $1^{+}_{1}\to2^{+}_{1}$ & 5.02 & $5.24$ \\
& $1^{+}_{1}\to2^{+}_{2}$ & 5.74 & $6.5$ \\
& $1^{+}_{1}\to2^{+}_{3}$ & 5.02 & $7.9$ \\
& $1^{+}_{1}\to2^{+}_{4}$ & 5.93 & $6.9$ \\
& $1^{+}_{1}\to2^{+}_{5}$ & 4.40 & $6.8$ \\
& $1^{+}_{1}\to2^{+}_{6}$ & 5.41 & $7.8$ \\
& $6^{+}_{1}\to6^{+}_{1}$ & 8.95 & $8.02\pm0.24$ \\
& $6^{+}_{1}\to5^{+}_{1}$ & 7.74 & $5.087\pm0.018$ \\
& $6^{+}_{1}\to5^{+}_{2}$ & 7.57 & $5.08\pm0.5$ \\
 \end{tabular}
 \end{ruledtabular}
 \end{center}
\end{table}

In addition to the transitions between 
the $0^+_1$ and $1^+_1$ states, 
there are $\ft$ data for the decays between states with 
spin other than $1^+$ and between higher-lying 
$0^+$ and $1^+$ states. 
To keep the discussion as simple as possible, 
let us focus on the $\beta$ decays 
that only involve the 
even-even $N=60$ isotones. 
The calculated and experimental \cite{data} 
$\ft$ values of the $\btm$ and EC 
decays are listed in Tables~\ref{tab:ft-btm} 
and \ref{tab:ft-ec} list, respectively. 
As one can see from Table~\ref{tab:ft-btm}, 
it is rather hard to draw a solid 
conclusion on 
the quality of the mapped IBM-2 framework 
for the description of the $\ft$ values 
for many different $\btm$ decays. 
Nevertheless, for many of the low-$Z$ nuclei, 
the present calculation gives smaller $\ft$ values 
than the experimental ones, suggesting 
that the assumption of considering only 
the single proton orbital ($\pi g_{9/2}$) 
may not be reasonable, and that the $\ft$ 
calculations are also influenced by 
the chosen parameters or forms of the 
IBFFM-2 Hamiltonian. 
The calculation, however, 
overestimates the measured $\ft$ values 
for higher-$Z$ nuclei.

Experimental data for the 
$\ft$ values for the EC decay are 
available for the $^{104}$Rh 
and $^{106}$Ag nuclei. 
Overall, the present calculation 
reproduces the $\ft$ data fairly well, 
especially for those nuclei that are not very 
far from the proton $Z=50$ major shell.

%
%
\begin{figure*}
\begin{center}
\includegraphics[width=.8\linewidth]{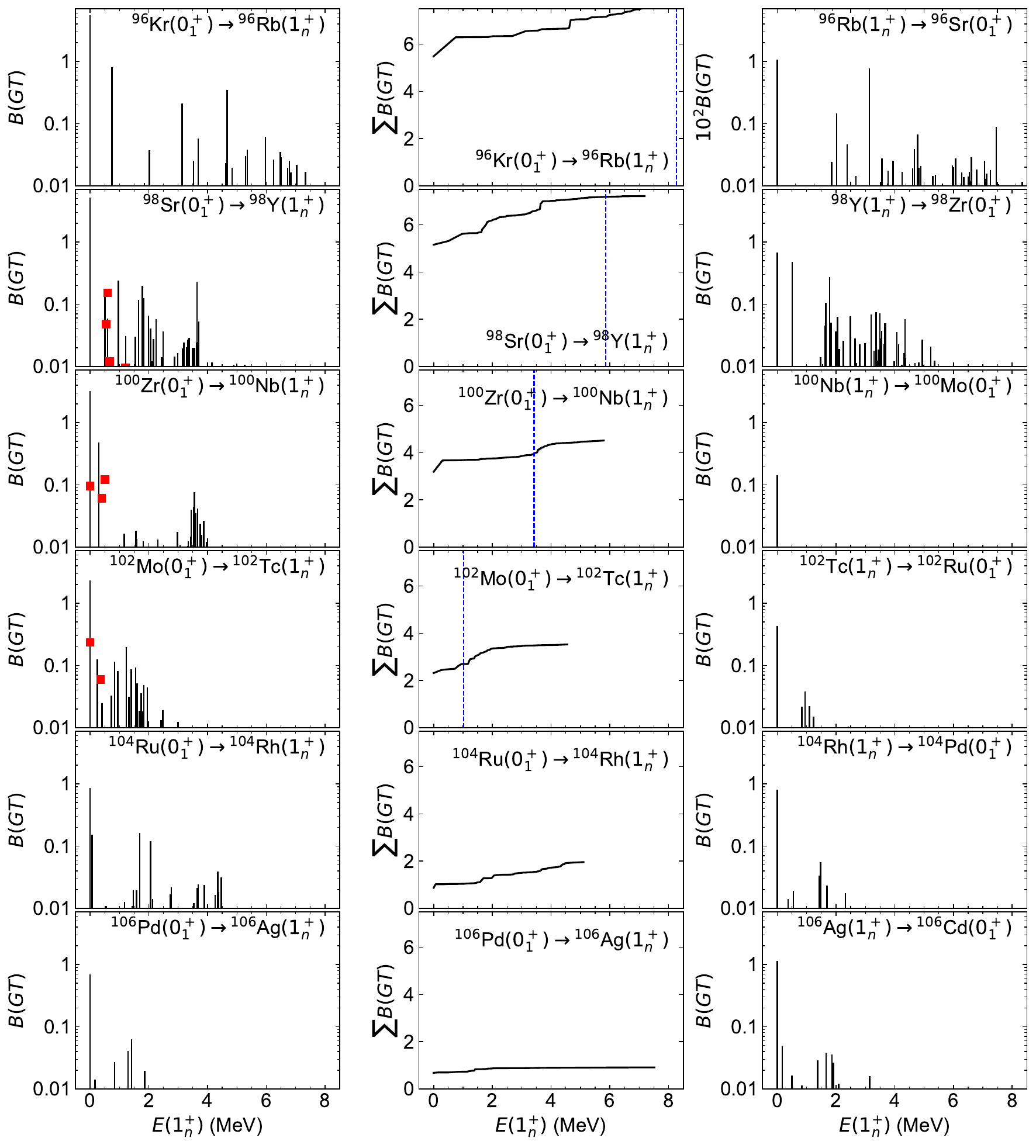}
\caption{$\bgtbeta$ strength distribution for the 
$\btm$ decay $0^+_1 \to 1^+_n$ of the 
even-even $N=60$ isotones in terms of 
the excitation energies of the $1^+_n$ 
states of the odd-odd final nuclei (left),  
their running sum (middle), and 
$\bgtbeta$ strength distribution for the 
$\btm$ decay $1^+_n \to 0^+_1$ 
of the odd-odd $N=59$ isotones (right). 
The experimental $B({\textnormal{GT}}; 0^+_1 \to 1^+)$ 
values extracted from the $\ft$ data 
\cite{data} are shown as red squares. 
The dashed vertical lines in the middle panels 
indicate experimental $Q_{\btm}$ values \cite{data}. 
Note that the 
$\bgtbeta$ values for the $^{96}$Rb$(1^+_n)\to^{96}$Sr$(0^+_1)$ 
are too small, and they are scaled with 
a factor of 100 so that they are more visible.
}
\label{fig:gt-dist-n60}
\end{center}
\end{figure*}

\subsection{GT strength distributions}

On the left-hand side of 
Fig.~\ref{fig:gt-dist-n60},  
distributions of the $B({\textnormal{GT}})$ transition strengths, 
$B({\textnormal{GT}}; 0^+_1 \to 1^+_1)=|\mgt(0^+_1 \to 1^+_1)|^2$, 
for the $\btm$ decay of the 
even-even $N=60$ isotones are shown 
as functions of the 
$1^+$ excitation energies 
$E(1^+_n)$ below 8 MeV. 
For the decays of $^{98}$Sr, $^{100}$Zr, and $^{102}$Mo, 
experimental $B({\textnormal{GT}})$ values are 
shown, which are obtained by using the 
corresponding $\ft$ data and the formula 
in Eq.~(\ref{eq:ft}). 
Error bars for the experimental $B({\textnormal{GT}})$ 
values are not shown, since they are 
smaller than the size of the symbols in the plot.

For each odd-odd nucleus, 
all the $1^+$ states resulting from the IBFFM-2 
and the corresponding GT transitions 
are here considered. 
For most of the even-even nuclei shown in 
the figure, the GT transition to 
the first excited $1^+_1$ is the strongest, 
while contributions from the decays to higher lying 
$1^+$ states become more minor with the 
increasing $1^+$ excitation energy. 
For the 
$^{104}$Ru$(0^+_1)\to^{104}$Rh$(1^+)$ 
decay, in particular, 
non-negligible amounts of the GT transitions 
are predicted within the excitation energies 
$E(1^+_n)$ from around 2 to 4 MeV. 
A similar degree of the fragmentation 
of the $\bgtbeta$ 
strength distributions are obtained for the 
$^{102}$Mo$(0^+_1)\to^{102}$Tc$(1^+)$ decay below 
$E(1^+_n)\approx 2$ MeV. 
The predicted GT strengths 
for the $^{98}$Sr, $^{100}$Zr, and $^{102}$Mo 
nuclei are, in some cases, larger by about 
an order of magnitude, but agree rather well 
with the observed systematic 
that the GT transitions to the low-lying $1^+$ 
states are dominant.

In addition, in the middle column of Fig.~\ref{fig:gt-dist-n60}, 
running sums of the GT strengths, i.e., 
$\sum_{n} B({\textnormal{GT}};0^+_1 \to 1^+_n)$, 
that are taken up to the excitation energy of 
the highest-lying $1^+$ state are shown. 
Experimental $Q_{\btm}$ values for the $^{96}$Kr, 
$^{98}$Sr, $^{100}$Zr, and $^{102}$Mo 
are also indicated in the figure. 
Note that, as $^{104}$Ru and $^{106}$Pd are stable 
nuclei, the corresponding $Q_{\btm}$ values 
are negative and thus not shown in the figure. 
For most of the considered $\btm$ decays, 
the sums appear to converge at low 
excitation energy, typically of $E(1^+)\leqslant 4$ MeV. 
Especially for the 
$^{98}$Sr$(0^+_1)\to^{98}$Y$(1^+)$ and 
$^{100}$Zr$(0^+_1)\to^{100}$Nb$(1^+)$ decays, 
the running sum more or less becomes 
constant before and around the measured 
$Q_{\btm}$ values, respectively. 
Notably for the $^{96}$Kr$(0^+_1)\to^{96}$Rb$(1^+)$ and 
$^{104}$Ru$(0^+_1)\to^{104}$Rh$(1^+)$ decays, 
the sums continue to increase up to the 
excitation energies of the highest-lying 
$1^+$ states obtained with the IBFFM-2. 
The GT decay 
rates are predicted to be remarkably 
large for the low-$Z$ 
isotones, i.e., $^{96}$Kr and $^{98}$Sr, 
with the corresponding running sums 
reaching 
$\sum_{n} B(\textnormal{GT};0^+_1 \to 1^+_n) \approx 7.5$. 
For higher-$Z$, 
stable nuclei, e.g., $^{106}$Pd, 
the final sum becomes smaller,  
$\sum_{n} B(\textnormal{GT};0^+_1 \to 1^+_n) < 1$. 
This conforms to the results shown 
in Fig.~\ref{fig:ft-even}, that the 
calculated $\ft$ values for the GT transition 
of the even-even nuclei to the lowest $1^+$ 
state of the 
odd-odd nuclei are particularly 
small for the low-$Z$ nuclei 
such as Kr and Sr ones.

On the right-hand side of 
Fig.~\ref{fig:gt-dist-n60}, 
the $\bgtbeta$ strength distributions, 
$B({\textnormal{GT}}:1^+_1 \to 0^+_1)=|\mgt(1^+_1 \to 0^+_1)|^2/3$, 
for the $\btm$ decays of the odd-odd $N=59$ 
into the even-even $N=58$ isotones 
are depicted 
in terms of the $E(1^+)$ excitation energy. 
In general, the GT strengths 
for the odd-odd isotones are  
predicted to be smaller, $<1.0$, 
than those for the even-even nuclei, 
which are larger than 1 
(see left column of Fig.~\ref{fig:gt-dist-n60}). 
As in the case of the even-even nuclei, 
contributions from the low-lying $1^+$ states 
are dominant also for the $\btm$ decays of 
odd-odd nuclei, such as 
$^{100}$Nb, $^{102}$Tc, 
$^{104}$Rh, and $^{106}$Ag. 
Of particular interest are the 
$^{96}$Rb$(1^+)\to^{96}$Sr$(0^+_1)$ and 
$^{98}$Y$(1^+)\to^{98}$Zr$(0^+_1)$ decays. 
For these decay processes, 
the $\bgtbeta$ strength distribution 
exhibits a substantial degree of fragmentation, 
and contributions from the nonyrast $1^+$ 
states are as significant as that 
from the lowest $1^+$ state. 
Note that, as for the $\bgtbeta$ rate of 
the $^{96}$Rb$(1^+)\to^{96}$Sr$(0^+_1)$ 
decay, the $\bgtbeta$ rates are negligibly 
small, so that they are scaled with 
a factor of 100 in the figure.

Summarizing the results shown in 
Fig.~\ref{fig:gt-dist-n60}, 
it appears that the GT transitions of 
the lowest or low-lying $1^+$ states make 
dominant roles in the running sums 
of the $\bgtbeta$ strengths 
for the $\btm$ decays of 
both the even-even and odd-odd nuclei, 
while the fragmentation tends to occur 
for those nuclei far from the 
proton $Z=50$ major shell closure. 
This observation seems compatible, 
to a good extent, to the single-state 
dominance \cite{griffiths1992,civitarese1998} 
or the low-lying state dominance \cite{moreno2008}
proposed especially 
for the studies of double-$\beta$ decay. 
The previous mapped IBM-2 calculation 
in Ref.~\cite{nomura2022bb} has also 
provided a similar conclusion on the 
$\tnbb$ decay NMEs of a number of candidate nuclei. 
The finding in the present calculation, that 
the GT transition strength is dominated 
by the low-lying $1^+$ states, sheds light 
upon the interpretation of the dominance in the 
GT strength, as it varies significantly 
with both the measurements and different 
theoretical approaches.

%
%
\begin{figure}
\begin{center}
\includegraphics[width=\linewidth]{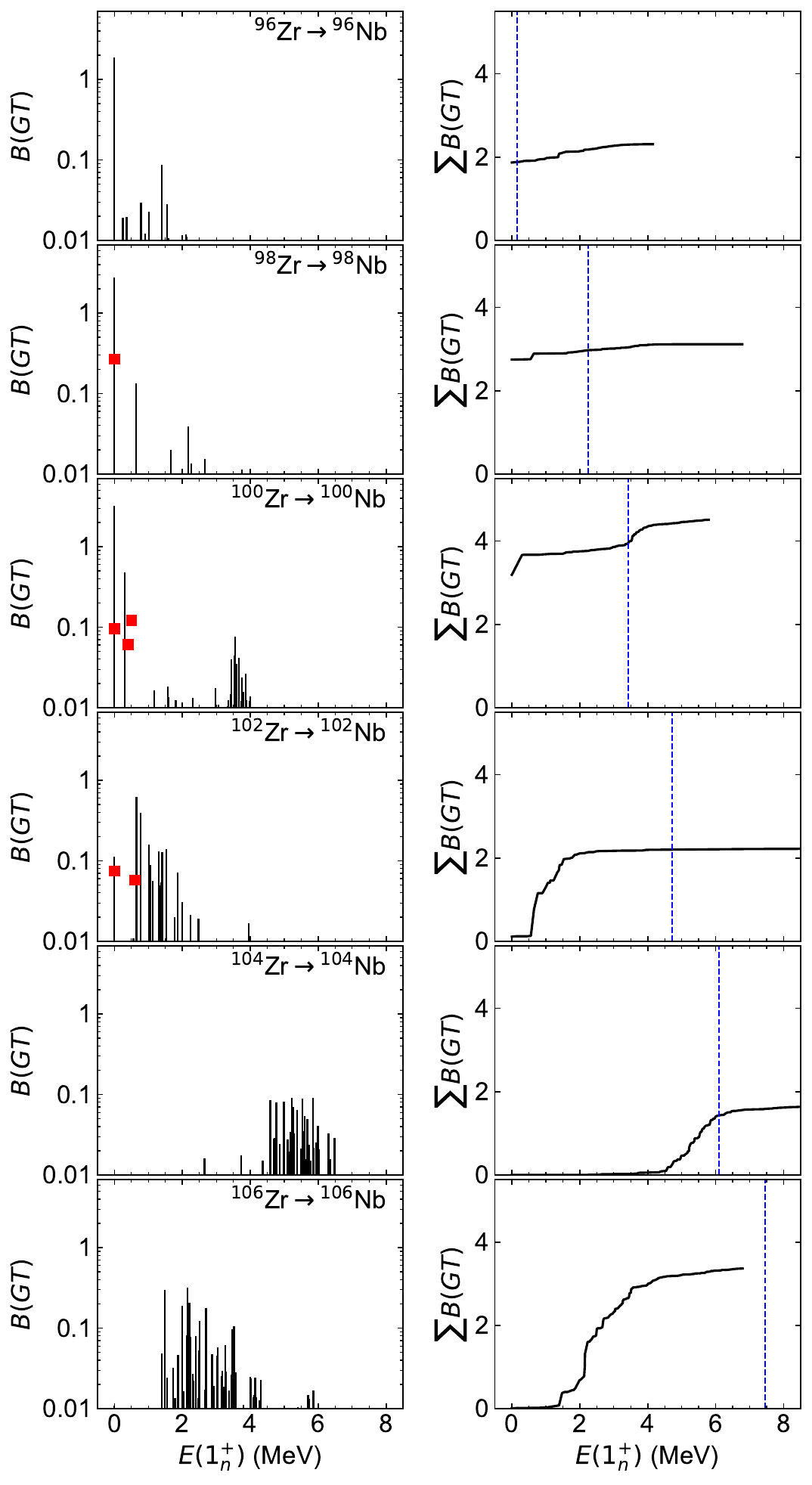}
\caption{
$\bgtbeta$ strength distribution for the 
$\btm$ decay $0^+_1 \to 1^+_n$ of the 
Zr isotopes in terms of 
the excitation energies $E(1^+_n)$ 
of the final Nb nuclei (left), and 
their running sums (right).}
\label{fig:gt-dist_zr-nb}
\end{center}
\end{figure}

It should be also worth investigating 
how the GT strength distribution changes 
along a given isotopic chain, particularly 
toward the neutron-rich region that is 
experimentally of much relevance. 
For that purpose, the $\bgtbeta$ strength 
distributions and their 
running sums for $\btm$ decays of 
the even-even Zr isotopes 
are shown in Fig.~\ref{fig:gt-dist_zr-nb}. 
For $^{96}$Zr, $^{98}$Zr, and $^{100}$Zr, 
the GT strength is almost solely accounted 
for by the transition to the first $1^+$ state, 
which is also considerably large in 
magnitude, 
$B(\textnormal{GT}; 0^+_1 \to 1^+_1) \approx 2-3$. 
However, one notices 
for the heavier Zr nuclei, i.e., 
$^{102}$Zr, $^{104}$Zr, and $^{106}$Zr, 
that the GT strength is fragmented 
to a great extent. 
Particularly for the 
$^{104}$Zr$(0^+_1)\to^{104}$Nb$(1^+)$ decay, 
the major contribution to the total GT strength 
comes from the transitions to the 
$1^+$ states that are at the 
excitation energy $E(1^+)$ of about 4-6 MeV, 
while the $B(\textnormal{GT};0^+_1 \to 1^+_1)$ 
transition does not play a significant role. 
For the lighter Zr nuclei, 
e.g., $^{96}$Zr and $^{98}$Zr, 
the sum, $\sum_n \bgtbeta$, 
appears to be converged at 
relatively low $E(1^+)$ excitation energies, 
$E(1^+)\approx$ 2 MeV, and 
the convergence is reached near the measured 
$Q_{\btm}$ values. 
Note that $^{96}$Zr is a candidate 
nucleus for the $\znbb$ decay. 
However, for the heavier Zr isotopes, 
$^{102}$Zr, $^{104}$Zr, and $^{106}$Zr, 
the convergence seems to occur at higher 
$E(1^+)$ excitation energies, e.g., $E(1^+) > 4$ MeV, 
that are larger or in the vicinity 
of the $Q_{\btm}$ values.

\begin{table*}
\caption{\label{tab:edf-para}
Strength parameters for the IBFFM-2 
Hamiltonian for the $^{106}$Rh nucleus employed 
in the present work (DD-PC1) and in the 
previous calculation that is based on the 
Gogny-D1M EDF \cite{D1M}, 
reported in Ref.~\cite{nomura2022beta-rh}. 
All the parameters, but for $\chi_\nu$ and $\chi_\pi$, 
are in MeV units.}
 \begin{center}
 \begin{ruledtabular}
  \begin{tabular}{lcccccccccccccc}
EDF 
& $\epsilon$ & $\kappa$ & $\chi_\nu$ & $\chi_\pi$ & $\kappa'$ 
& $\Gamma_\nu$ & $\Lambda_\nu$ & $A_\nu$ 
& $\Gamma_\pi$ & $\Lambda_\pi$ & $A_\pi$ 
& $\vd$ & $\vt$ \\
\hline
DD-PC1 & 0.18 & $-0.364$ & $-0.5$ & $-0.38$ & 0.032 &
0.3 & 0 & 0 & 0.3 & 1.2 & 0 & $-0.08$ & 0.05 \\
D1M & 0.85 & $-0.301$ & $-0.86$ & $-0.6$ & 0 &
0.7 & 0.7 & 0 & 0.6 & 0.6 & 0 & $-0.20$ & 0.1 \\
  \end{tabular}
 \end{ruledtabular}
 \end{center}
\end{table*}

\begin{table}
\caption{\label{tab:edf-spe}
The quasiparticle energies, $\tilde\epsilon_{\jr}$ (in MeV), 
and occupation probabilities, $v^2_{\jr}$, for the 
single-particle orbitals used in the mapped IBM-2 
calculations employing the DD-PC1 and D1M EDFs.}
 \begin{center}
 \begin{ruledtabular}
  \begin{tabular}{llccccc}
EDF &
& $\nu 3s_{1/2}$ & $\nu 2d_{3/2}$ & $\nu 2d_{5/2}$ & $\nu 1g_{7/2}$ & $\pi 1g_{9/2}$ \\
\hline
\multirow{2}{*}{DD-PC1} 
& $\tilde\epsilon_{\jr}$
& $1.883$ & $2.148$ & $1.674$ & $1.760$ 
& $1.463$ \\
& $v^2_{\jr}$
& $0.158$ & $0.147$ & $0.705$ & $0.693$ 
& $0.522$ \\
\multirow{2}{*}{D1M} 
& $\tilde\epsilon_{\jr}$
& $3.706$ & $4.446$ & $1.863$ & $3.214$ 
& $1.166$ \\
& $v^2_{\jr}$
& $0.333$ & $0.179$ & $0.843$ & $0.496$ 
& $0.516$ \\
  \end{tabular}
 \end{ruledtabular}
 \end{center}
\end{table}

\subsection{Sensitivity to the choice of EDF\label{sec:edf}}

As a possible source of the theoretical uncertainty, 
in this section sensitivity of the $\ft$ predictions 
to the choice of the EDF is explored. 
The $^{106}$Rh$\to^{106}$Pd decay is 
considered as an illustration, since this decay was also 
studied within the mapped IBM-2/IBFFM-2 framework 
in Ref.~\cite{nomura2022beta-rh}, 
where the D1M interaction \cite{D1M} of 
the Gogny EDF was employed as the microscopic 
input for the IBM-2 and IBFFM-2. 
The present calculation with the DD-PC1 
EDF generally gives larger $\ft$ values 
than that based on the Gogny-D1M EDF. 
For example, 
the Gogny-D1M EDF calculation of 
Ref.~\cite{nomura2022beta-rh} provided the 
$\ft$ value of the $1^+_1 \to 0^+_1$ decay 
to be 3.31, while the present value is 
4.46 (see Table~\ref{tab:ft-btm}), which is 
larger than the former 
by an order of magnitude in $ft$ values.

Table~\ref{tab:edf-para} 
compares the IBFFM-2 parameters for $^{106}$Rh 
obtained with the DD-PC1 and Gogny-D1M EDFs. 
Differences in the parameters for the IBM-2 
Hamiltonian are due to the different nature 
of the PESs provided by the two functionals. 
In both cases, the $(\beta,\gamma)$ PES for the 
even-even core nucleus, $^{106}$Pd, shows a 
weak prolate 
deformation typically at $\beta \approx 0.2$ 
(see Fig.~\ref{fig:pes}, 
and Fig.~2 of Ref.~\cite{nomura2022beta-rh}). 
The DD-PC1 PES, however, exhibits a 
steeper potential valley than the D1M counterpart. 
This is reflected in the differences in 
the derived parameters, e.g., 
$\epsilon$ and $\kappa$, 
shown in Table~\ref{tab:edf-para}. 
Furthermore, in Table~\ref{tab:edf-spe} 
the quasiparticle energies, $\tilde\epsilon_{\jr}$, 
and occupation probabilities, $v^2_{\jr}$, 
for the odd neutron and odd proton, 
which are used in determining 
the IBFFM-2 interaction strength and 
$\beta$-decay operators, are compared between 
the mapped IBM-2 calculations based on the 
DD-PC1 and Gogny-D1M EDFs. 
These single-particle properties 
are crucial to determine the boson-fermion 
and fermion-fermion interactions, 
but are at variance with the EDFs.

%
%
\begin{figure}
\begin{center}
\includegraphics[width=.8\linewidth]{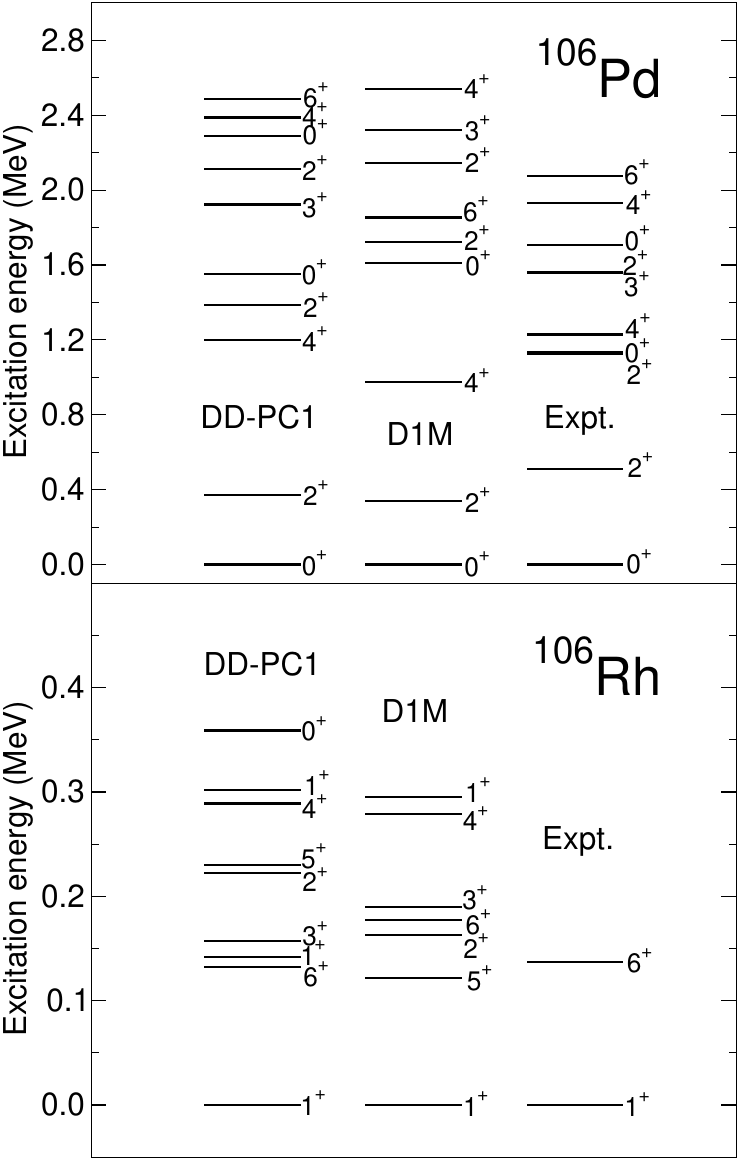}
\caption{Low-energy spectra for the 
$^{106}$Pd and $^{106}$Rh nuclei computed with 
the microscopic inputs provided from the 
DD-PC1 and Gogny-D1M EDFs. The experimental data 
are taken from Ref.~\cite{data}, and the 
results for the Gogny-D1M EDF are exploited 
from Ref.~\cite{nomura2022beta-rh}.}
\label{fig:pd106-rh106_edf}
\end{center}
\end{figure}

Figure~\ref{fig:pd106-rh106_edf} compares 
the energy spectra for the $^{106}$Pd 
and $^{106}$Rh nuclei between the calculations 
based on the DD-PC1 and Gogny-D1M EDFs. 
In general, the mapped IBM-2 calculation with 
the DD-PC1 EDF gives a better description 
of the energy spectra for $^{106}$Pd than 
that employing the D1M EDF. 
The D1M-IBM-2 energy spectrum exhibits 
a rotational feature and is rather stretched 
in energy with increasing angular momentum, 
as compared to the DD-PC1 result. 
The difference arises mainly from the parameters 
$\chi_\nu$ and $\chi_{\pi}$, which are 
responsible for the $\gamma$ softness. 
Especially when the absolute value 
of the average, 
$\chi=|\chi_\nu + \chi_{\pi}|/2$, is chosen to be 
large, then it generally leads to 
a rotational-like level structure. 
One actually sees from Table~\ref{tab:edf-para} 
that both the parameters $\chi_\nu$ and $\chi_{\pi}$ 
employed in the previous IBM-2 study \cite{nomura2022beta-rh} 
with the D1M EDF are larger in magnitude 
than those in the present calculation using 
the DD-PC1 EDF.

Similarly, for the odd-odd nucleus, 
$^{106}$Rh, the energy spectra 
obtained from the two EDFs are 
rather different, that is, 
the calculation employing the 
DD-PC1 EDF reproduces better the $6^+_1$ 
energy level, while in the D1M energy spectrum 
the $2^+_1$ and $5^+_1$ levels come below 
the $6^+_1$ level. 
In addition, the IBFFM-2 calculation 
with the DD-PC1 EDF gives 
nonyrast $1^+$ energy levels that are lower 
than in the case of the D1M EDF. 
In the DD-PC1 EDF calculation, 
the structure of the IBFFM-2 
$1^+_1$ state for $^{106}$Rh is described 
by the neutron-proton pair configurations 
$[\nu g_{7/2}\otimes \pi g_{9/2}]^{(J=1)}$ 
and 
$[\nu g_{7/2}\otimes \pi g_{9/2}]^{(J=3)}$, 
which share, respectively, 
52 \% and 43 \% of the wave function. 
In the IBFFM-2 calculation of Ref.~\cite{nomura2022beta-rh} 
using the Gogny-D1M EDF, however, 
only the $[\nu g_{7/2}\otimes \pi g_{9/2}]^{(J=1)}$ 
pair component plays a predominant role (73 \%)
to determine the $1^+_1$ wave function, 
whereas there are numerous minor 
contributions from other pair configurations. 
This difference more or less explains 
the discrepancies in the 
$\ft$ values between the two EDFs, 
since in the Gogny-D1M case 
matrix elements of the terms in the 
GT operator that contain the pair 
$[\nu g_{7/2}\otimes \pi g_{9/2}]^{(J=1)}$
are quite large, but those of other terms 
cancel each other to a greater extent 
than in the case of the DD-PC1 EDF.

As it is also evident from the comparisons in 
Tables~\ref{tab:edf-para}, \ref{tab:edf-spe}, 
and Fig.~\ref{fig:pd106-rh106_edf}, 
the parameters for the IBFFM-2 and 
single-particle properties depend largely 
on the choice of the EDF, 
which then lead to the differences 
in the predicted excitation spectra and 
$\ft$ values. 
It remains, however, an open issue to identify 
which parameters are most responsible for 
the different $\ft$ value predictions 
in the employed theoretical approach.

\section{Summary and conclusions\label{sec:summary}}

The low-energy structure 
and $\beta$-decay properties of the 
neutron-rich even-mass nuclei around 
$N=60$ that are currently under extensive 
investigations have been studied 
within the theoretical framework 
of the EDF-to-IBM mapping. 
The IBM-2 Hamiltonian describing 
the even-even core nuclei, 
and particle-boson interactions 
have been determined by using 
the results of the triaxial quadrupole 
constrained SCMF calculations within the 
RHB model with the DD-PC1 functional  
and the separable pairing force. 
The resulting wave functions for the 
initial and final nuclei, obtained 
from the IBM-2 and IBFFM-2, 
have been used to compute the GT 
transition strengths, where the corresponding 
operators have been constructed without 
introducing adjustable parameters.

The calculated $(\beta,\gamma)$ PESs 
for the $N=60$ even-even 
isotones suggest for most cases notably 
$\gamma$-soft shapes that vary substantially 
with $Z$. 
Spectroscopic calculations 
for the even-even 
Kr, Sr, Zr, Mo, Ru, Pd, and Cd isotopes 
within the mapped IBM-2 have produced evolution 
of the low-lying energy spectra and $B(E2)$ 
rates as functions of $N$, and rapid 
changes of these observables around $N=60$ 
along the Sr, Zr, and Mo isotopic chains. 
The excitation spectra of low-spin states 
of the neighboring odd-odd nuclei 
have been shown to exhibit certain variation 
with $N$, reflecting the shape transitions 
that occur in the even-even core nuclei.

The mapped IBM-2 and IBFFM-2 calculation 
has provided the $\ft$ values 
for the $\btm$ decays of the $0^+_1$ 
state of the even-even nuclei into 
the $1^+_1$ state of the odd-odd nuclei 
that are systematically lower than 
the experimental values for 
the lower-$Z$ isotopes (i.e., Kr, Sr and Zr) 
and mostly for $N \leqslant 60$. 
The too small $\ft$ values in these 
nuclear systems indicate a need for 
a substantial degree of quenching for the 
$\ga$ factor, which amounts to 
an order of magnitude in some cases. 
For those nuclei with $N>60$ and higher $Z$, 
however, the calculated $\ft$ values 
exhibit a sharp increase with $N$, and 
especially for Mo isotopes the calculation 
overestimates the data. 
The necessity of introducing effective $\ga$ 
factors would then imply deficiencies 
of the theoretical framework that 
arise from various model assumptions 
and approximations, including the 
particular choice of the EDF providing 
microscopic input to the IBM-2 and IBFFM-2, 
the form of the corresponding Hamiltonians, 
and the restricted single-particle space. 
The sensitivity of the results to the 
choice of the nuclear EDF has been explored, 
specifically, by comparing the calculations 
that employ the relativistic DD-PC1 and 
nonrelativistic Gogny-D1M EDFs as the 
microscopic inputs to the IBM-2 and IBFFM-2.

The simultaneous calculation of the low-energy 
nuclear structure and $\beta$ decay 
will be useful for improving the quality 
of the employed theoretical method 
in describing spectroscopic properties 
of individual nuclei even more accurately, 
and will provide implications for 
studies of other fundamental nuclear processes 
including the $\znbb$ decay.

\bibliography{refs}

\end{document}